\documentclass[twocolumn]{aa}
\usepackage{graphicx}
\usepackage{txfonts}
\usepackage{comment}
\usepackage{float}
\usepackage{subfig}
\usepackage[dvipsnames]{xcolor}
\usepackage[utf8]{inputenc}
\usepackage[T1]{fontenc}
\usepackage{amsfonts}
\usepackage[colorlinks=true,allcolors=blue]{hyperref}
\usepackage{mathrsfs}

\begin{document} 
\title{Grain alignment and dust evolution physics with polarisation (GRADE-POL)}
\subtitle{I. Dust polarisation modelling for isolated starless cores}
\author{Le Ngoc Tram\inst{1,2}\thanks{Corresponding author: Le Ngoc Tram \newline \email{nle@strw.leidenuniv.nl, nle@mpifr-bonn.mpg.de}},  
Thiem Hoang\inst{3,4},
Alex Lazarian\inst{5},
Daniel Seifried\inst{6},
B-G Andersson\inst{7},
Thushara G.S. Pillai\inst{8},
Bao Truong\inst{3},
Pham Ngoc Diep\inst{9,10},
Lapo Fanciullo\inst{11}
}
\institute{$^{1}$ Leiden Observatory, Leiden University, PO Box 9513, 2300 RA Leiden, The Netherlands \\
$^{2}$Max-Planck-Institut für Radioastronomie, Auf dem Hügel 69, 53121, Bonn, Germany \\
$^{3}$ Korea Astronomy and Space Science Institute, Daejeon 34055, Republic of Korea \\
$^{4}$ Department of Astronomy and Space Science, University of Science and Technology, 217 Gajeong-ro, Yuseong-gu, Daejeon, 34113, Republic of Korea
\\
$^{5}$ Department of Astronomy, University of Wisconsin-Madison, Madison, WI, 53706, USA \\
$^{6}$ I. Physikalisches Institut, Universität zu Köln, Zülpicher Str. 77, D-50937 Köln, Germany \\
$^{7}$ McDonald Observatory, University of Texas at Austin, 2515 Speedway Boulevard, Austin, TX 78712, USA \\
$^{8}$ Haystack Observatory, Massachusetts Institute of Technology, 99 Millstone Road, Westford, MA~01886, USA \\
$^{9}$ Department of Astrophysics, Vietnam National Space Center, Vietnam Academy of Science and Technology, 18 Hoang Quoc Viet, Hanoi, Vietnam \\
$^{10}$ Graduate University of Science and Technology, Vietnam Academy of Science and Technology, 18 Hoang Quoc Viet, Hanoi, Vietnam\\
$^{11}$ National Chung Hsing University, 145 Xingda Rd., South Dist., Taichung City 402, Taiwan \\
} 
\titlerunning{Grain Alignment and Evolution in Starless Cores}
\authorrunning{Tram et al., 2025}
\abstract
{The polarisation of light induced by aligned interstellar dust serves as a significant tool in investigating cosmic magnetic fields and dust properties, while posing a challenge in characterising the polarisation of the cosmic microwave background and other sources. To establish dust polarisation as a reliable tool, the physics of the grain alignment process must be studied thoroughly. The magnetically enhanced radiative torque (MRAT) alignment is the only mechanism that can induce highly efficient alignment of grains with magnetic fields required by polarisation observations of the diffuse interstellar medium. Here, we aim to test the MRAT mechanism in starless cores using the multi-wavelength polarisation from optical to submillimetre. Our numerical modelling of dust polarisation using the MRAT theory demonstrates that the alignment efficiency of starlight polarisation ($p_{\rm ext}/A_{\rm V}$) and the degree of thermal dust polarisation ($p_{\rm em}$) first decrease slowly with increasing visual extinction ($A_{\rm V}$) and then fall steeply as $\propto A^{-1}_{\rm V}$ at large $A_{\rm V}$ due to the loss of grain alignment, which explains the phenomenon known as polarisation holes. Visual extinction at the transition from shallow to steep slope ($A^{\rm loss}_{\rm V}$) increases with maximum grain size. By applying physical profiles suitable for a starless core, 109 in the Pipe nebula (Pipe-109), our model successfully reproduces the existing observations of starlight polarisation in the R band ($0.65\,\mu$m) and the H band ($1.65\,\mu$m), as well as emission polarisation in the submillimetre ($870\,\mu$m). Successful modelling of observational data requires perfect alignment of large grains, which serves as evidence for the MRAT mechanism, and an increased maximum grain size with higher elongation at higher $A_{\rm V}$. The latter reveals the first evidence for a new model of anisotropic grain growth induced by magnetic grain alignment. This paper introduces the framework for probing the fundamental physics of grain alignment and dust evolution using multi-wavelength dust polarisation (GRADE-POL), and it is the first of our GRADE-POL series.}

\maketitle
\section{Introduction}\label{sec:intro}
The polarisation of starlight and dust emission caused by alignment of dust grains (hereafter dust polarisation) is commonly used to detect magnetic fields in the interstellar medium (ISM) of our Galaxy (e.g. \citealt{2023ASPC..534..193P}) and external galaxies, both nearby (e.g. \citealt{2023ApJ...952....4B}) and far away (e.g. \citealt{2023Natur.621..483G}). It also presents a challenge as foreground contamination in the characterisation of the cosmic microwave background (e.g. \citealt{2020A&A...641A...4P}), exoplanetary atmospheres (e.g. \citealt{2021A&A...647A..21V}), and other sources. Hence, a thorough understanding of the physical processes behind dust polarisation and grain alignment is necessary.

After detections of the extinction (starlight) polarisation (\citealt{1949Sci...109..166H,1949Sci...109..165H}) and emission (thermal dust) polarisation (\citealt{1982MNRAS.200.1169C}), numerous studies have been carried out to establish a theory for the physics of grain alignment. The alignment by radiative torques (RATs),  initially proposed by \cite{1976Ap&SS..43..291D}, has been widely accepted as a leading theory of grain alignment. This RAT idea was supported by numerical simulations in \citet{DW1996}. Subsequently, the rotation of dust grains exposed to a laser beam was observed in a laboratory experiment \citep{2004ApJ...614..781A}.
A simple analytical model for helical grains presented in \citet{2007MNRAS.378..910L} and numerical calculations in \cite{2008MNRAS.388..117H} enabled efficient exploration of the parameter space for the dynamics of dust grains subjected to radiation. These analytical and numerical studies showed that the fraction of grains that can be aligned at an attractor point with high angular momentum $J$ (called high$-J$ attractors) by RATs strongly varies with the angle between the radiation and interstellar magnetic field directions. This angle dependence was also reported from observations in \cite{2011A&A...534A..19A}. Comprehensive studies on RAT alignment for a large sample of grain shapes are presented in \citet{2019ApJ...878...96H,2021ApJ...913...63H}. For the ensemble of shapes considered in \citealt{2021ApJ...913...63H}, on average, only about $20-40\%$ of the grains can be aligned at high$-J$ attractors by RATs for a typical interstellar radiation field (ISRF; see their Figure 5).

For grains with embedded iron inclusions, the effect of magnetic relaxation cannot be ignored in the consideration of RAT alignment. The theoretical study in \cite{2008ApJ...676L..25L} shows that the enhancement in magnetic susceptibility due to iron inclusions could increase the fraction of grains aligned at high-$J$ attractors. For example, for a small magnetic inclusion, about 70$\%$ of the grains can align at high$-J$ attractors, and a considerable iron inclusion can enable $100\%$ of grains to align at high-$J$ attractor for the same ISRF (see \citealt{2016ApJ...831..159H} and Figure 10 in \citealt{2021ApJ...913...63H}). In particular, numerical calculations in \cite{2016ApJ...831..159H} show that grains can be perfectly aligned by a magnetically enhanced RAT (also known as MRAT) mechanism. The classical RAT theory has been shown to qualitatively reproduce dust polarisation observations from diffuse to molecular clouds (MCs; see review in \citealt{2015ARA&A..53..501A}). However, a quantitative comparison between theoretical modelling and Planck observational data reveals the need for highly efficient grain alignment in the diffuse ISM \citep{Hensley.2023} and in MCs \citep{Reissl.2020}, which requires the MRAT mechanism. In particular, the high polarisation of up to $30-40\%$ observed in dense protostellar environments by the Atacama Large Millimeter Array (ALMA) provides strong support for the MRAT mechanism (\citealt{2023MNRAS.520.3788G,2024MNRAS.530..984G,2024ApJ...970..114T,2024arXiv240710079C}). In this paper, we seek evidence of MRAT in dense starless cores, which are intermediate structures between MCs and protostellar cores.

The continuous development of the RAT alignment theory \citep{2019NatAs...3..766H, 2020ApJ...891...38H} demonstrates that in the presence of a strong radiation field, large dust grains rotating at high$-J$ attractors can achieve an exceptionally high angular velocity. These grains spontaneously disrupt into smaller fragments once the centrifugal force within the grains exceeds the binding force that holds them together. This effect of rotational disruption is termed radiative torque disruption (RAT-D). Laboratory experiments show extremely rapid rotation of objects due to laser irradiation (e.g. \citealt{2018PhRvL.121c3602R,2018PhRvL.121c3603A}), and the disruptive impact of rotation is also observed in simulations (\citealt{2023arXiv230112889R}) and laboratory experiments (\citealt{2019EL....12745004H}). The combination of RAT alignment with RAT-D mechanisms constitutes the RAT paradigm, which accounts for a wider range of dust polarisation towards star-forming regions (see the review in \citealt{2022FrASS...9.3927T}). Thus, the RAT paradigm is a foundational framework to describe grain alignment and grain disruption driven by radiation.

To extend the application of the RAT paradigm in various astrophysical environments, we developed a physical (forward) model to predict the degree of polarisation at multiple wavelengths for both starlight polarisation, $p_{\rm ext}$ and thermal dust polarisation, $p_{\rm em}$. This physical model is based on the alignment degree predicted by the RAT paradigm, but it assumes a uniform magnetic field, neglects the depolarisation owing to fluctuations of this field, and excludes the radiative transfer process. The model is coded in \textsc{python} and described in \cite{2020ApJ...896...44L} and \cite{2021ApJ...906..115T}. We name this code \textsc{DustPOL\_py}\footnote{Github: \href{https://github.com/lengoctram/DustPOL_py}{https://github.com/lengoctram/DustPOL\_py}}. Despite its simplicity, \textsc{DustPOL\_py} reproduces the relation of $p_{\rm em}$ with dust temperature (a proxy for radiation intensity) in star-forming regions, both when averaging over the cloud (\citealt{2020ApJ...896...44L,2021ApJ...906..115T,2021ApJ...908..159T}) and on a map (pixel-by-pixel; \citealt{2024arXiv240316857B}). This model also reproduces the spectrum of thermal dust polarisation, from 54 to 850$\,\mu$m in the Orion Molecular Cloud 1 (OMC-1) (\citealt{2024arXiv240317088T}). In these early investigations, the model is treated as zero-dimensional and includes local physical parameters such as the gas volume density and the dust temperature, which are derived from observations. Furthermore, the model's ability to predict starlight polarisation is not thoroughly utilised.

Another longstanding problem in dust astrophysics is how dust grains grow in the ISM and in dense MCs. It is well known that grains can grow by gas accretion and grain-grain collisions, but the detailed physics is unclear. Traditionally, grain growth is assumed to be isotropic because of the random motion of gas and grains, which results only in an increase in grain size (see, e.g. \citealt{Hirashita.2013}); however, the evolution of grain shape during grain growth has not yet been studied in detail. Therefore, some authors suggest that grain growth makes large grains more spherical and thus reduces dust polarisation efficiency, which might explain the polarisation hole \citep{2000prpl.conf..247W,gupta2024magnetic}. However, this suggestion conflicts with the high degree of polarisation of starlight and dust emission in dense protostellar cores observed by ALMA, where grain growth is most efficient. A recent theoretical study in \cite{2022ApJ...928..102H} reveals two new effects of grain alignment on grain growth. First, the growth of aligned grains through gas accretion becomes anisotropic due to the alignment of grains with the magnetic field. This results in an increase in grain elongation during the grain growth process. Second, grain growth by coagulation via collisions between aligned grains results in the formation of composite grains with elongation increasing with grain growth. Polarisation is a unique tool to constrain both grain size and shape, as shown in the pioneering study for the ISM by \cite{draine2024sensitivity}. Nevertheless, the shape of dust grains and their effect on dust polarisation in dense clouds, where grain growth is most efficient, are not well studied. Compared to the prolate shape, oblate grains show a better fit to the observed mid-infrared polarisation towards Orion Becklin–Neugebauer (\citealt{1989MNRAS.236..919A}) and the Chamaeleon I cloud (\citealt{2010ApJ...720.1045A,2011A&A...534A..19A}). However, \cite{2021ApJ...909...94D} showed that both prolate and oblate grains can reproduce mid-infrared polarisation towards two lines of sight in the diffuse ISM presented in \cite{2002osp..conf...85W}. Recently, \cite{Reissl.2024} have shown that the fast rotation of grains and the resulting centrifugal force can cause them to have oblate shapes, with elongation increasing with the rotation rate. Since RATs cause grain alignment and suprathermal rotation, they are also important for the evolution of grains in size (growth and disruption) and shapes (oblate and elongation).

In this work, our objective is to constrain the physics of grain alignment and dust evolution using dust polarisation (GRADE-POL). Towards this goal, we improve our \textsc{DustPOL-py} code and apply it to isolated, starless, spherical cores. Starless cores are ideal targets for our study because of the high densities at which grain growth is expected to occur, and grain alignment remains significant. Numerous starless cores have been observed at multiple wavelengths, from starlight polarisation in optical and near-infrared (near-IR) to thermal dust polarisation in the submillimetre (submm) range. These observations usually report a drop in $p_{\rm ext}/A_{\rm V}$ and/or $p_{\rm em}$ from the cloud surface to the core centre as a function of the total intensity ($I$) or visual extinction ($A_{\rm V}$), following a power law ($p_{\rm ext}/A_{\rm V} \sim A_{\rm V}^{-\alpha}$ or $p_{\rm em}\sim A^{-\alpha}_{\rm V}$) (see, e.g. \citealt{1995ApJ...455L.171G,2008ApJ...674..304W,2014A&A...569L...1A,2014ApJS..213...13H}). In some cases, this slope, $\alpha$, approaches $1$ in close proximity to the core centres (\citealt{2015AJ....149...31J}), which is known as the polarisation hole.

Earlier studies have attempted to interpret observed dust polarisation in starless cores, using the original classical RAT theory from \cite{2007MNRAS.378..910L}, for starlight polarisation (\citealt{2008ApJ...674..304W,2015MNRAS.448.1178H}) and thermal dust polarisation (\citealt{2005ApJ...631..361C,2007ApJ...663.1055B}). Most recently, based on the RAT paradigm, the theoretical work in \cite{2021ApJ...908..218H}, with a simple radiative transfer consideration, explains that the slope of $p_{\rm em} \sim A_{\rm V}^{-1}$ could result from the loss of aligned grains and predicts that the $p_{\rm em}-A_{\rm V}$ relation could be used to investigate grain growth in starless cores. In this paper, we simultaneously model the multi-wavelength polarisation of starlight and thermal dust emission using the state-of-the-art RAT paradigm and compare it with observational data.

In our previous research, we have explored three scenarios for dust grain population: separate silicate and carbonaceous particles, and mixed compositions. Specifically, using multiple wavelengths of thermal dust polarisation in \cite{2024arXiv240317088T}, we demonstrate that a combined composition of carbonaceous and silicate (bonded together) is necessary to understand observations towards the OMC-1. This mixture leads to a flat polarisation spectrum in the submm range because of the slightly different temperatures of the silicate and carbonaceous grains. This characteristic is similar to the new model of composite cosmic dust (\textsc{Astrodust}) introduced by \cite{2023ApJ...948...55H}, which effectively reproduces the flat spectrum observed by {\it Planck} and the BlastPOL balloon. Composite interstellar dust was first proposed in \cite{1989ApJ...341..808M,1989IAUS..135..239T}. In addition, numerical simulations in \cite{2023arXiv230112889R} show that a rotating grain is likely to have an oblate shape. In this work, we incorporate and consider only the oblate \textsc{astrodust} composition.

This paper is structured as follows. We summarise the fundamental mathematical formalism for the physical properties of starless cores and alignment by RATs used in \textsc{DustPOL-py} in Section \ref{sec:model_formulation}. Section \ref{sec:numerical_results} presents the numerical results. We discuss our findings in Section \ref{sec:discussion}, in which we model the multi-wavelength polarisation and compare it to observations for the Pipe-109 starless core at optical, near-IR, and submm wavelengths. Conclusions are given in Section \ref{sec:conclusion}.

\section{Model formalism} \label{sec:model_formulation}
Here, we first review the foundational formulations and physical parameters that underlie our model, including the astrophysical object and dust alignment. For the dust model, we consider the optical constants of the new \textsc{Astrodust} (\citealt{2023ApJ...948...55H}).

\subsection{Numerical setup}
\begin{figure}
    \centering
    \includegraphics[width=0.33\textwidth]{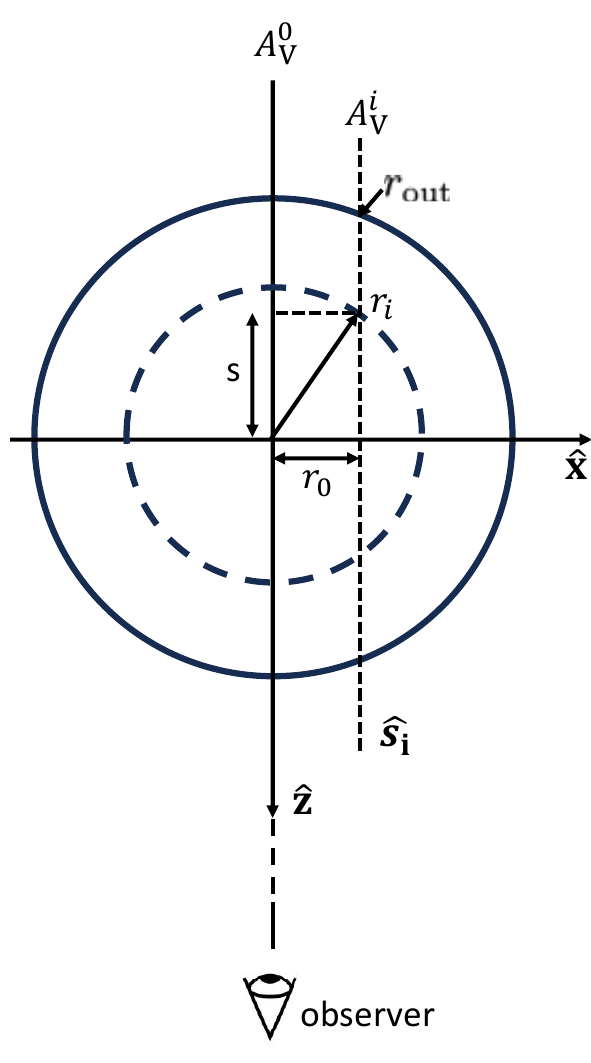}
    \caption{Coordinate system of the starless core on the {\it(x,z)}-plane with the ${\hat{z}}$ direction towards the observer. The position in the plane of the sky is defined by $r_{0}$. Each line of sight, $\hat{s}$, corresponds to a distinct visual extinction, $A_{\rm V}$, calculated from the gas volume density profile along this direction. At a location $r_{i}$ on the sight line, the physical parameters (such as gas density, temperature, and radiation field) are assumed to be uniform on the dashed circle.}
    \label{fig:sketch}
\end{figure}

This work focuses on starless cores (or dense clouds), which are assumed to have a spherical geometry with radius $r_{\rm out}$ (see Figure \ref{fig:sketch}). The line of sight follows the $\hat{z}$ direction in Cartesian coordinates.

Calculations were carried out on the {\it $\hat{x}\hat{z}$ plane} with $y=0$ (the largest circular slice). We sampled 102 values of $r_{0}$ from 0 to $r_{\rm out}$, and 200 values of $s$ from $-r_{\rm out}$ to $r_{\rm rout}$. The samplings were in linear space. For a given line of sight, $\hat{s}_{i}$, with a projected distance to the centre of the core of $r_{0}$, the radial distance along this line of sight was $r_{i} = \pm \sqrt{s^{2}+r^{2}_{0}}$ with $|s|$ varying from $-r_{\rm out}$ to $r_{\rm out}$. Due to the assumption of spherical symmetry of the starless core, the calculations were not executed for $r_{i}>r_{\rm out}$. 

For each line of sight at position $r_{0}$, the visual extinction $A^{i}_{\rm V}$ observed by the observer was calculated as
\begin{equation}\label{eq:Av_obs}
    A_{\rm V}(r_{0}) = 1.086\tau_{\rm 0.55\,\mu m}(r_{0})
\end{equation}
with the optical depth given by
\begin{equation}
        \tau_{0.55\,\rm \mu m}(r_{0}) =\int_{s}n_{\rm H} ds\left(\int^{a_{\rm max}}_{a_{\rm min}}\pi a^{2}Q_{\rm ext, 0.55\,\mu m}(a)\frac{1}{n_{\rm H}}\frac{dn}{da}da\right).
\end{equation}
Here, $Q_{\rm ext}$ is the extinction efficiency of the dust grain and $dn/da$ is the grain size distribution with minimum size ($a_{\rm min}$) and maximum size ($a_{\rm max}$). For a given grain size distribution, visual extinction is sensitive to the chosen maximum grain size; for example, a smaller $a_{\rm max}$ results in a higher $A_{\rm V}$. We did not model the evolution of grain size within the cores.

\subsection{Physical profiles}
This section outlines the parametric expressions for the physical profiles applied in our model. We refer to \cite{2021ApJ...908..218H} for the original and detailed derivations.
\subsubsection{Gas density and attenuation}
We considered a typical density model of starless cores with a flat central peak and a decreasing gas density outward. Let $n_{0}$ be the gas density in the central region and $r$ the radial distance from the centre. The density profile is described by
\begin{equation}
    n_{\rm H}(r)=\left\{
    \begin{array}{l l}
        n_{\rm 0} & \quad {\rm ~for~} r\le r_{\rm flat},\\
        n_{\rm 0}\left(\frac{r}{r_{\rm flat}}\right)^{-\alpha} & \quad {\rm ~for~} r>r_{\rm flat},
    \end{array}\right.
\end{equation}
where $r$ extends to $r_{\rm out}$, with $n_{0}$, $r_{\rm out}$, $r_{\rm flat}$, and $\alpha$ treated as parameters within our model. We considered only the particular case of $\alpha=2$, consistent with \cite{2001ApJ...547..317W}. The other three parameters were adjusted based on observational constraints, e.g. gas volume density or visual extinction. The case of $\alpha=0$ corresponds to a uniform molecular cloud.

The gas column density ($N_{\rm H}$), integrated from the centre to the position $r$ in the envelope along the radial distance, and the corresponding visual extinction ($A^{\rm outward}_{V}$) were
\begin{equation}
    \begin{split}
        N_{\rm H}(r) &= \int^{r}_{0} n_{\rm H}(r')dr' ~~~ {\rm (cm^{-2}),}\\
        A^{\rm outward}_{\rm V} &= \left(\frac{N_{\rm H}}{5.8\times 10^{21}\,\rm cm^{-2}}\right)R_{\rm V} ~~~ {\rm (mag),} 
    \end{split}
\end{equation}
where $R_{\rm V}$ denotes the total-to-selective extinction ratio. We adopted $R_{\rm V}=4$ to account for the larger grains relative to the typical interstellar medium. The visual extinction ($A^{\rm ext}_{\rm V}$), measured from the envelope to the centre, were expressed as
\begin{equation}
    \begin{split}
    A^{\rm ext}_{\rm V} &= A^{\rm outward}_{\rm V}(r\gg r_{\rm flat}) - A^{\rm outward}_{\rm V}(r) \\
    &= 10.3\left(\frac{n_{0}}{10^{8}\,\rm cm^{-3}}\right)\left(\frac{r_{\rm flat}}{10\,\rm au}\right)\left(\frac{R_{\rm V}}{4}\right) \\ 
    & \times \left\{
    \begin{array}{l l}
        \left(\frac{\alpha}{\alpha-1} - \frac{r}{r_{\rm flat}}\right) & \quad {\rm ~for~} r\le r_{\rm flat},\\
        \frac{1}{\alpha-1}\left(\frac{r}{r_{\rm flat}}\right)^{1-\alpha} & \quad {\rm ~for~} r>r_{\rm flat},
    \end{array}\right.
    \end{split}
\end{equation}
It should be noted that this visual extinction characterises the extinction from the surface to the position $r$ along the radial direction and differs from the line of sight $A_{\rm V}$ in Equation \ref{eq:Av_obs}.

\subsubsection{Radiation field and grain temperature}
Due to the lack of embedded internal radiation sources within a starless core, the local radiation field inside the starless core was just an attenuated interstellar radiation field (ISRF). Let $u_{\rm rad}$ be the radiation energy density and $U$ be the strength of the local radiation field, defined as $U=u_{\rm rad}/u_{\rm MMP}$ where $u_{\rm MMP}= 8.64\times 10^{-13}{\rm erg} {\rm cm}^{-3}$ is the energy density of the local radiation field in the solar neighbourhood \citep{1983A&A...128..212M}. 
Following numerical calculations in \cite{2021ApJ...908..218H}, who fitted empirical functions to the results of radiative transfer calculations, the strength of the attenuated ISRF at a visual extinction $A_{V}^{\rm ext}$ inside a starless core was parameterised as
\begin{equation}
    U(A^{\rm ext}_{\rm V}) = \frac{\int^{\infty}_{0}u_{\lambda}(A^{\rm ext}_{\rm v})d\lambda}{u_{\rm MMP}} = \frac{U_{0}}{1+0.42\times \left(A^{\rm ext}_{\rm V}\right)^{1.22}}.
    \label{eq:U_AV}
\end{equation}
Here, $u_{\lambda}$ is the specific radiation energy density in units of $\rm erg\,cm^{-4}$, and $U_{0}$ is the radiation strength measured at the cloud surface, which is assumed to be the same as that of the diffuse ISRF. For a typical ISRF in the solar neighbourhood \citep{1983A&A...128..212M}, $U_{0}=1$, but a recent study suggests $U_{0}= 1.5$ \citep{Bianchi.2024}. 

The corresponding mean wavelength of the ISRF was
\begin{equation}
    \bar{\lambda}(A^{\rm ext}_{\rm V}) = \frac{\int_{0}^{\infty} \lambda u_{\lambda}(A^{\rm ext}_{\rm V})d\lambda}{\int_{0}^{\infty} u_{\lambda}(A^{\rm ext}_{\rm V})d\lambda} = \bar{\lambda}_{0}\left[1+0.27\times \left(A^{\rm ext}_{\rm V}\right)^{0.76}\right],
\end{equation}
where $\bar{\lambda}_{0}$ is the mean wavelength of the ISRF at the cloud surface, which is typically $\bar{\lambda}_{0}=1.2\,\mu$m.

We assumed that gas and dust were in thermal equilibrium, i.e. $T_{\rm gas}=T_{\rm dust}$. Therefore, one can determine the gas and dust temperatures using the radiation strength given by Equation \ref{eq:U_AV} and the formula from \cite{2011piim.book.....D}
\begin{equation}
    T_{\rm d}(A^{\rm ext}_{\rm V}) = T_{\rm gas}(A^{\rm ext}_{\rm V}) = 16.4\,{\rm K}\times \left(\frac{a}{0.1\,\mu m}\right)^{-1/15}\left[U\left(A^{\rm ext}_{\rm V}\right)\right]^{1/6}.
\end{equation}

\subsubsection{Grain size distribution}
We used the well-known MRN-like grain size distribution with $dn/da \sim C a^{\beta}$ (with $\beta$ being a negative number). The normalisation factor, $C$, was estimated by mass conversion as (\citealt{2020ApJ...893..138T,2021ApJ...906..115T})\footnote{There was a typo in \cite{2020ApJ...893..138T} missing the factor $n_{\rm H}$ in the writing, but the calculation took this factor into account.}
\begin{equation}
    \begin{split}
        C & = \frac{(4+\beta)M_{\rm d/g}m_{\rm gas}}{\frac{4}{3}\pi \rho_{\rm dust}n_{\rm H}(a^{4+\beta}_{\rm max}-a^{4+\beta}_{\rm min})} & \quad {\rm ~for~~~} \beta \neq -4, \\
        C & = \frac{M_{\rm d/g}m_{\rm gas}}{\frac{4}{3}\pi \rho_{\rm dust}n_{\rm H}(\ln a_{\rm max}-\ln a_{\rm min})} & \quad {\rm ~for~~~} \beta=-4,
    \end{split}
\end{equation}
where $M_{\rm d/g}$ is the dust-to-gas mass ratio ($M_{\rm d/g}=0.01$ for a typical ISM), $\rho_{\rm dust}$ is the volume density of the dust mass ($\rho_{\rm dust}=2.71\,\rm g\,cm^{-3}$ for \textsc{astrodust} with 20\% porosity (\citealt{2023ApJ...948...55H}), and $m_{\rm gas}=2.8n({\rm H}_{2})m_{\rm H}\simeq 1.4n_{\rm H}m_{\rm H}$ is the mean molecular gas mass per hydrogen atom\footnote{If the cloud is fully molecular, one gets $n({\rm H}_{2})=n_{\rm H}/2$}. Note that this normalisation factor establishes the value of the line-of-sight extinction $A_{\rm V}$, but cancels out when considering the starlight polarisation to extinction ($p_{\rm ext}/A_{\rm V}$ -- mostly used in observations), as well as the thermal dust polarisation $p_{\rm em}$. In contrast, the power law of the distribution exerts a significant influence.

\subsection{Grain alignment by the radiative torque paradigm}
The physics of grain alignment by radiative torques has been described in detail in numerous works. In this section, we restate the principles of this theory and the formulations for our model. For more details, we refer to the most recent reviews in \cite{2015ARA&A..53..501A,2022FrASS...9.3927T} and studies in \cite{2021ApJ...908...12L,2022AJ....164..248H}. Moreover, with the adoption of the ISRF and the presence of dense gas in starless cores, RAT-D could not take place, leading us to disregard the formulation of this rotational disruption.
\subsubsection{Radiative torques}
The average radiative torque induced by the interaction of an irregular grain (with an effective size $a$) and an anisotropic radiation field (with isotropic degree $\gamma$) is
\begin{equation}
    \bar{\Gamma}_{\rm RAT} = \pi a^{2} \gamma u_{\rm rad}\left(\frac{\bar{\lambda}}{2\pi}\right)\bar{Q}_{\Gamma},
\end{equation}
where $u_{\rm rad}= Uu_{\rm MMP}$ is the energy density of the local radiation field, and
$\bar{Q}_{\Gamma}$ is the RAT efficiency given by
\begin{equation}\label{eq:Q_gamma}
    \bar{Q}_{\Gamma}=\left\{
    \begin{array}{l l}
        2\left(\frac{\bar{\lambda}}{a}\right)^{-2.7}   & \quad \text{for } a\leq \frac{\bar{\lambda}}{1.8},\\
        0.4  & \quad \text{for } a>\frac{\bar{\lambda}}{1.8}.
    \end{array}\right.
\end{equation}
The RAT efficiency power index of $-2.7$ and the transition at $\bar{\lambda}/1.8$ differ slightly from the slope of $-3$ and the transition at $\bar{\lambda}/2.7$ in \cite{2021ApJ...908..218H}. However, this discrepancy in the torques is very small, as shown in their Figure 1.

\subsubsection{Angular velocity and grain alignment size}
Radiative torques (RATs) can spin up and irregular grain to a finite angular velocity ($\omega_{\rm RAT}$). The maximum angular velocity that an irregular grain is gained by RAT is given by
\begin{equation}
    \omega_{\rm RAT} = \frac{\bar{\Gamma}_{\rm RAT}\tau_{\rm damp}}{I_{a}},
\end{equation}
where $I_{a}=8\pi\rho a^{5}/15$ is the grain moment of inertia, with $\rho$ being the grain mass density, and $\tau_{\rm damp}$ is the rotational damping time that characterises the damping caused by gas-grain collisions, followed by evaporation and infrared emission.

The rotational damping timescale is given by
\begin{equation} \label{eq:tau_damp}
    \begin{split}
        \tau_{\rm damp} \simeq 8.3\times 10^{3}&\left(\frac{a}{0.1\,\rm \mu m}\right)\left(\frac{\rho}{3\,\rm g\,cm^{-3}}\right) \\ 
        &\times \left(\frac{10^{3}\,\rm cm^{-3}}{n_{\rm H}}\right)\left(\frac{10\,\rm K}{T_{\rm gas}}\right)^{1/2} \left(\frac{1}{1+F_{\rm IR}}\right) ~~~ {\rm yr},
    \end{split}
\end{equation}
where $F_{\rm IR}$ is the ratio of damping timescales between gas collisions and infrared emission, which is approximately given by
\begin{equation}
    F_{\rm IR} = 0.038\times U^{2/3}\left(\frac{0.1\,\rm \mu m}{a}\right)\left(\frac{10^{3}\,\rm cm^{-3}}{n_{\rm H}}\right)\left(\frac{10\,\rm K}{T_{\rm gas}}\right)^{1/2} ~~~{.}
\end{equation}
When $F_{\rm IR} \ll 1$ (for instance, when $U^{2/3}/n_{\rm H} \ll 1$), the damping timescale of the grain is influenced by the density and temperature of the gas. Conversely, if $F_{\rm IR} \gg 1$ (for example, $U^{2/3}/n_{\rm H} \gg 1$), the damping timescale $\tau_{\rm damp} \sim n^{-1}_{\rm H}T^{-1/2}_{\rm gas}F^{-1}_{\rm IR}$ becomes independent of the gas density and temperature. In starless cores, the former scenario applies.

The grain alignment remains effective if its angular velocity is three times greater than the thermal angular velocity ($(k_{\rm B}T_{\rm gas}/I_{a})^{1/2}$) (see \citealt{2008MNRAS.388..117H}). Therefore, the minimum grain size that can be effectively aligned (referred to as the alignment size, $a_{\rm align}$) is the solution of the following equation: 
\begin{equation}
    \omega_{\rm RAT}(a\equiv a_{\rm align}) = 3\times \left(\frac{15k_{\rm B}T_{\rm gas}}{8\pi \rho a_{\rm align}^{5}}\right)^{1/2} ~~~{.}
\end{equation}

\subsubsection{Alignment function}
The alignment of grains with the ambient magnetic field can be characterised by two processes: internal alignment (where the grain's minor axis aligns with its angular momentum, $J$; see e.g. \citealt{1979ApJ...231..417S}) and external alignment (where the angular momentum aligns with a preferred axis, which could be the ambient magnetic field, radiation field, or gas flow; see \citealt{2022FrASS...9.3927T} for a review). The efficiency of internal alignment is described by $Q_{\rm X}$, while external alignment is described by $Q_{\rm J}$. The net grain alignment efficiency is quantified by the Rayleigh reduction factor of $R(a)=\langle Q_{X}Q_{J}\rangle$ where the brackets denote the average over an ensemble of grains of the given size $a$ (see \citealt{2015MNRAS.448.1178H}). Since grains aligned at high-J attractors have perfect internal alignment due to suprathermal rotation, i.e. $Q_{X}=1$. Therefore, we can approximate $R\simeq f_{\rm high-J}$. 

The MRAT theory predicts the dependence of $f_{\rm high}$ on the local gas density and the magnetic properties of the grain, with $f_{\rm high-J}=1$ for superparamagnetic grains \citep{2008ApJ...676L..25L,2016ApJ...831..159H}. Numerical calculations of grain alignment using MRAT theory \cite{2016ApJ...831..159H} show that the degree of alignment tends to increase smoothly with grain size, consistent with results from inverse modelling of starlight polarisation (\citealt{HoangLazMartin.2014}). Therefore, the alignment function can be parameterised as
\begin{equation}
    f(a) = f_{\rm max}\left[1-e^{-(0.5a/a_{\rm align})^{3}} \right],
\end{equation}
which implies $f(a)=f_{\rm max}$ for $a\gg a_{\rm align}$ and $f(a)\rightarrow 0$ for $a\ll a_{\rm align}$. The case of $f_{\rm max}=1$ is referred to as a perfect alignment of large grains by RATsaa.

\subsection{Degree of dust polarisation}

\subsubsection{Starlight polarisation}
The degree of starlight polarisation for a population of dust, along a given line of sight in the optically thin case is (see, e.g. \citealt{2015MNRAS.448.1178H})
\begin{equation} \label{eq:pext}
    p_{\rm ext}(A_{\rm V}) = 100\times \int_{s} n_{\rm H}ds \left(\int^{a_{\rm max}}_{a_{\rm min}}\pi a^{2}Q^{\rm pol}_{\rm ext}F_{\rm turb}\sin^{2}\psi f(a)\frac{1}{n_{\rm H}}\frac{dn}{da}da\right) ~~~\%,
\end{equation}
where $\psi$ is the inclination angle between the magnetic field orientation and the line-of--sight, $F_{\rm turb}$ is the turbulence factor characterising the fluctuation of the magnetic field, and $Q_{\rm ext}^{\rm pol}$ is the extinction polarisation coefficient. This efficiency is determined by the residual of the extinction efficiencies in the two components where the electric field is parallel and perpendicular to the grain rotation axis ($\boldsymbol{a}$), expressed as $0.5\left[Q_{\rm ext}(\boldsymbol{E}\parallel \boldsymbol{a})-Q_{\rm ext}(\boldsymbol{E}\perp \boldsymbol{a})\right]$. These components are taken from the \textsc{astrodust} database\footnote{\href{http://arks.princeton.edu/ark:/88435/dsp01qb98mj541}{http://arks.princeton.edu/ark:/88435/dsp01qb98mj541}}. Here, $F_{\rm turb}=0.5[3\langle \cos^{2}(\Delta \theta)\rangle -1]$, where $\Delta \theta$ denotes the angular deviation between the local magnetic field and the average field (see \citealt{2024arXiv240714896T}). Since local and average fields are difficult to disentangle, $F_{\rm turb}$ is generally difficult to define. However, when the cloud is sub-Alfvénic, this factor can be estimated via the angle dispersion of the magnetic fields, $\delta \theta$, as $F_{\rm turb}\simeq 1-1.5(\delta \theta)^{2}$. The expression $F_{\rm turb}\times \sin^{2}\psi$ in Equation \ref{eq:pext} determines the influence of magnetic field changes, both in 3D orientations and fluctuations. For clarity, we assume $F_{\rm turb}=1$ and $\psi=90^{o}$ in subsequent sections, while discussing their effects specially in the Pipe-109 case study in Section \ref{sec:Bfield_variation}.

\begin{table}[!ht]
    \centering
    \caption{Input parameters}
    \begin{tabular}{c|c|l}
        \hline \hline
         Parameter & Value & Description \\
         \hline
         $n_{0}$       & $5\times 10^{5}\,\rm cm^{-3}$ & Gas density at the core\\
         $\alpha$      & $-2$ & Slope of gas density\\
         $r_{\rm flat}$& $0.02\,\rm pc$ & Radius of uniform gas density\\
         $r_{\rm out}$ & $0.6\,$pc & Outer radius of the cloud\\
         $U_{0}$       & 1 & Ambient radiatio field (aISRF)\\
         $\gamma$      & 0.3 & Anisotropic degree of aISRF\\
         $\bar{\lambda_{0}}$ & 1.2$\,\mu$m & Mean-wavelength of aISRF\\
         $T_{\rm gas,0}$ & 16.4$\,$K & Gas temperature at the surface\\
         $\beta$ & $-3.5$ & Power-index of size distribution\\
         $f_{\rm max}$ & 1.0 & Maximum alignment efficiency\\
         $\psi$ & $90^{\circ}$ & Inclination angle of B-field\\
         $F_{\rm turb}$ & 1.0 & Turbulence factor of B-field\\
         \hline \hline
    \end{tabular}
    \begin{flushleft}
    \tablefoot{
   For demonstration purposes, $F_{\rm turb}$ is assigned a value of 1.0, representing a uniform magnetic field. It is recommended to determine the appropriate values tailored to each specific target. In this study, for Pipe-109, we computed $F_{\rm turb}$ as 0.98, 0.96, and 0.91 for the R-, H-bands, and submm wavelengths respectively. The values of $n_{0}$, $r_{\rm flat}$ and $r_{\rm out}$ are for the Pipe-109 (see Section \ref{sec:pipe-109}), and the value of $\gamma$ is used as a constant.
   }
    \end{flushleft}
    \label{tab:parameters}
\end{table}
\begin{figure}
    \centering
    \includegraphics[width=0.5\textwidth]{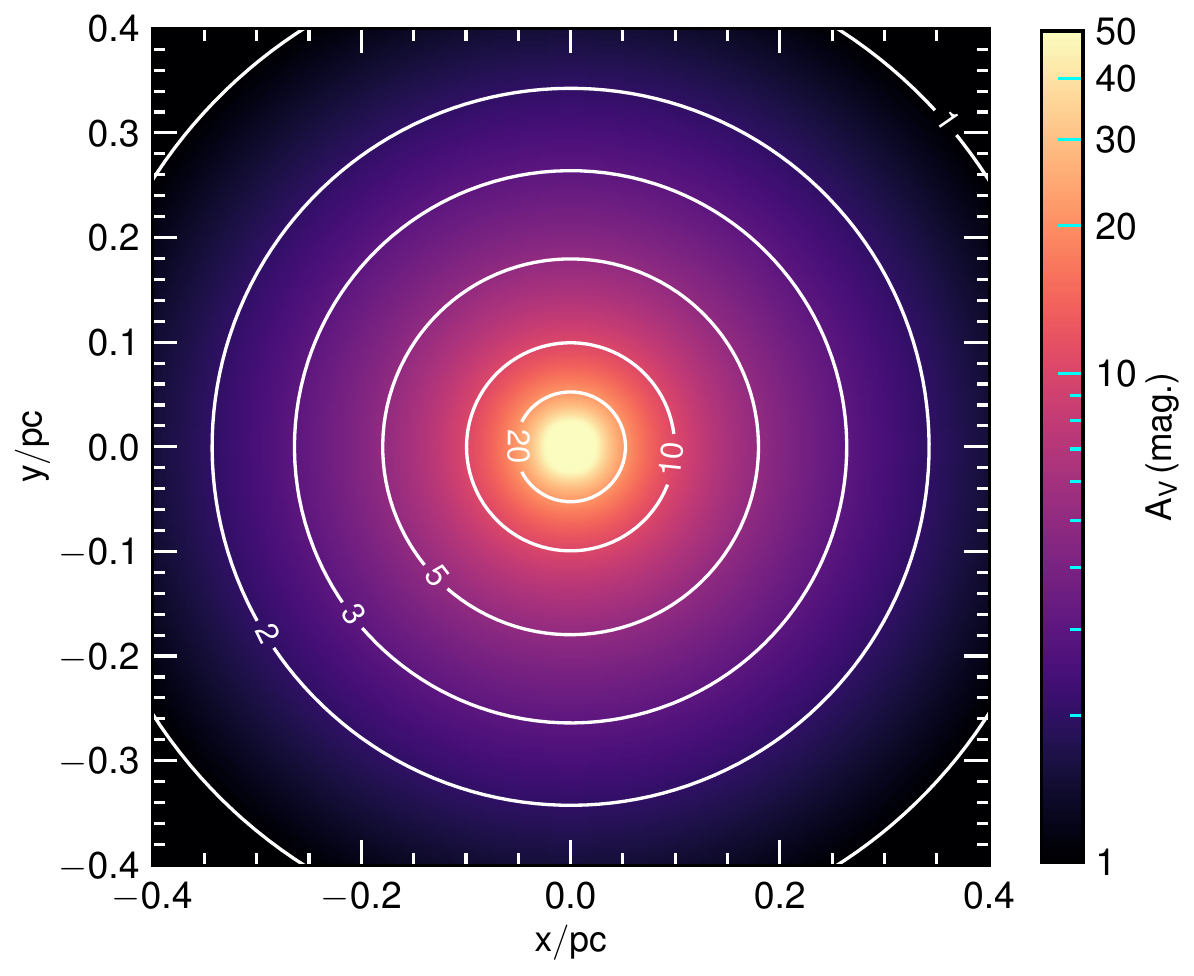}
    \includegraphics[width=0.5\textwidth]{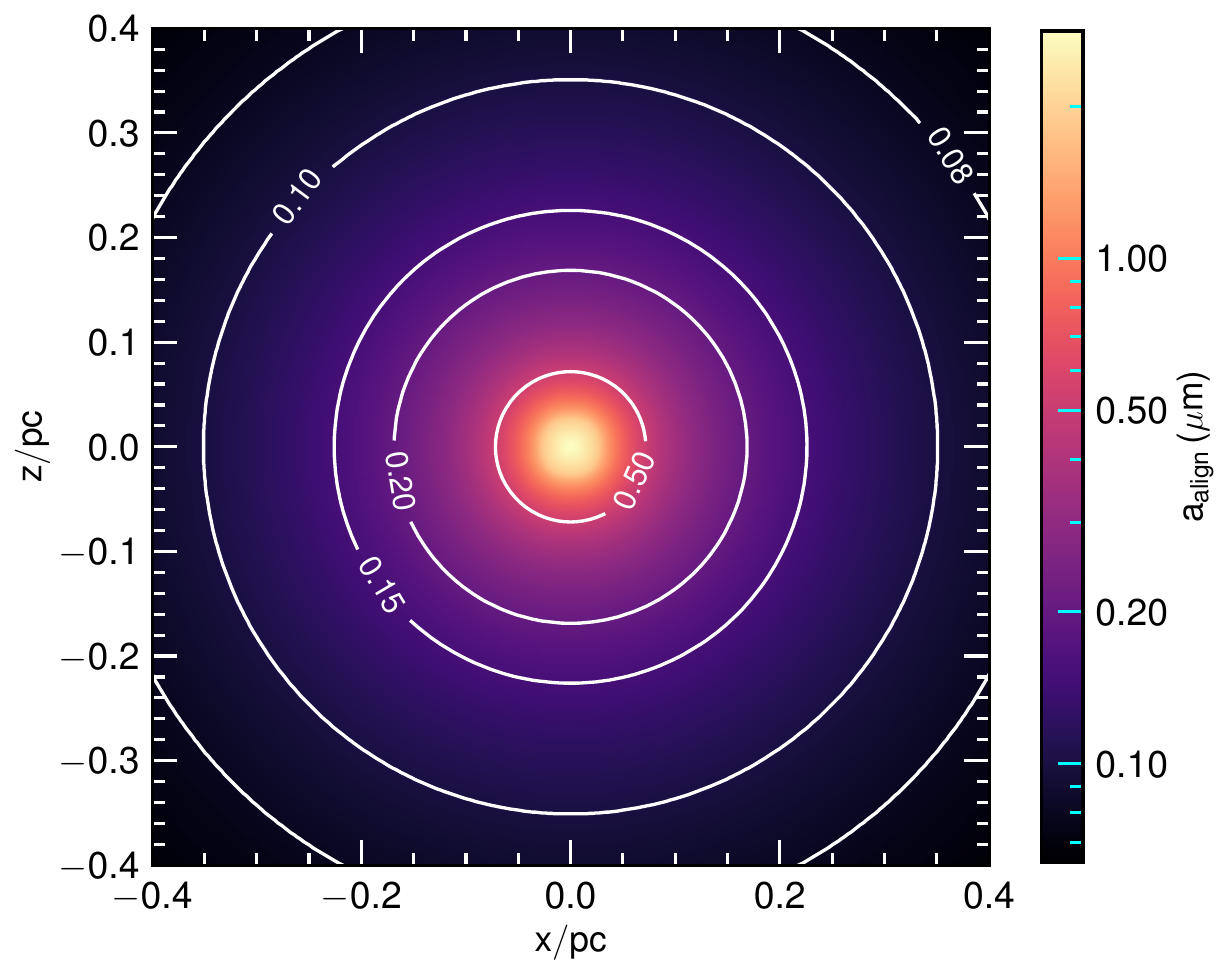}
    \caption{{\bf (Top panel)}: Full map on the plane of the sky ({\it $\hat{x}\hat{y}$ plane}) of the line-of-sight visual extinction. {\bf (Bottom panel)}: Alignment size along the line of sight ({\it $\hat{x}\hat{z}$ plane}). We adopt $n_{0}=5\times 10^{5}\,\rm cm^{-3}$, $r_{\rm flat}=0.024\,\rm pc$, $\alpha=2$, $U_{0}=1$, $\gamma=0.3$, oblate grains (axial ratio of 1.4) with $a_{\rm max}=1\,\mu$m, $\beta=-3.5$, and $f_{\rm max}=1$. This setup mimics the scale of the Pipe-109 starless core, whose data will be used later in this work.}
    \label{fig:phys_prop}
\end{figure}

\begin{figure*}
    \centering
    \includegraphics[width=0.48\textwidth]{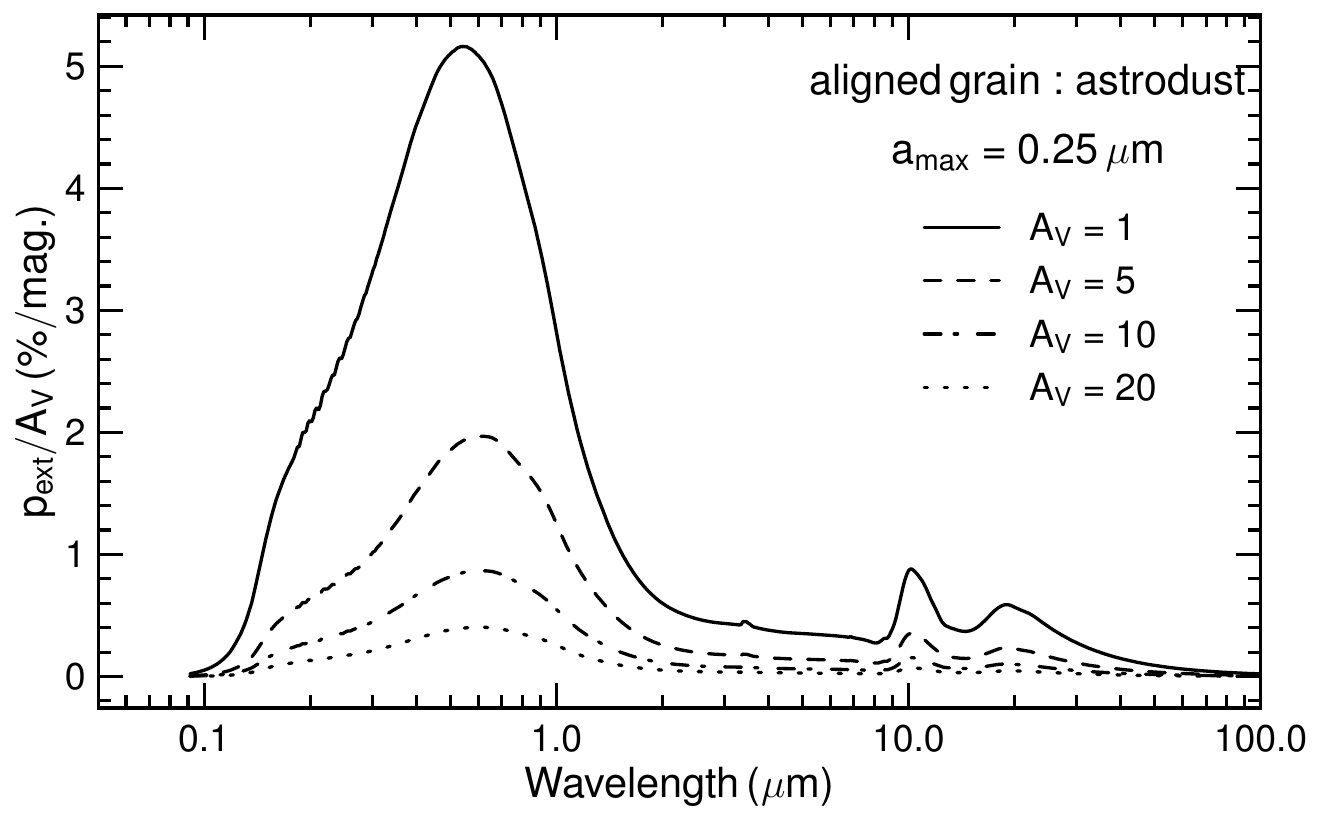}    
    \includegraphics[width=0.48\textwidth]{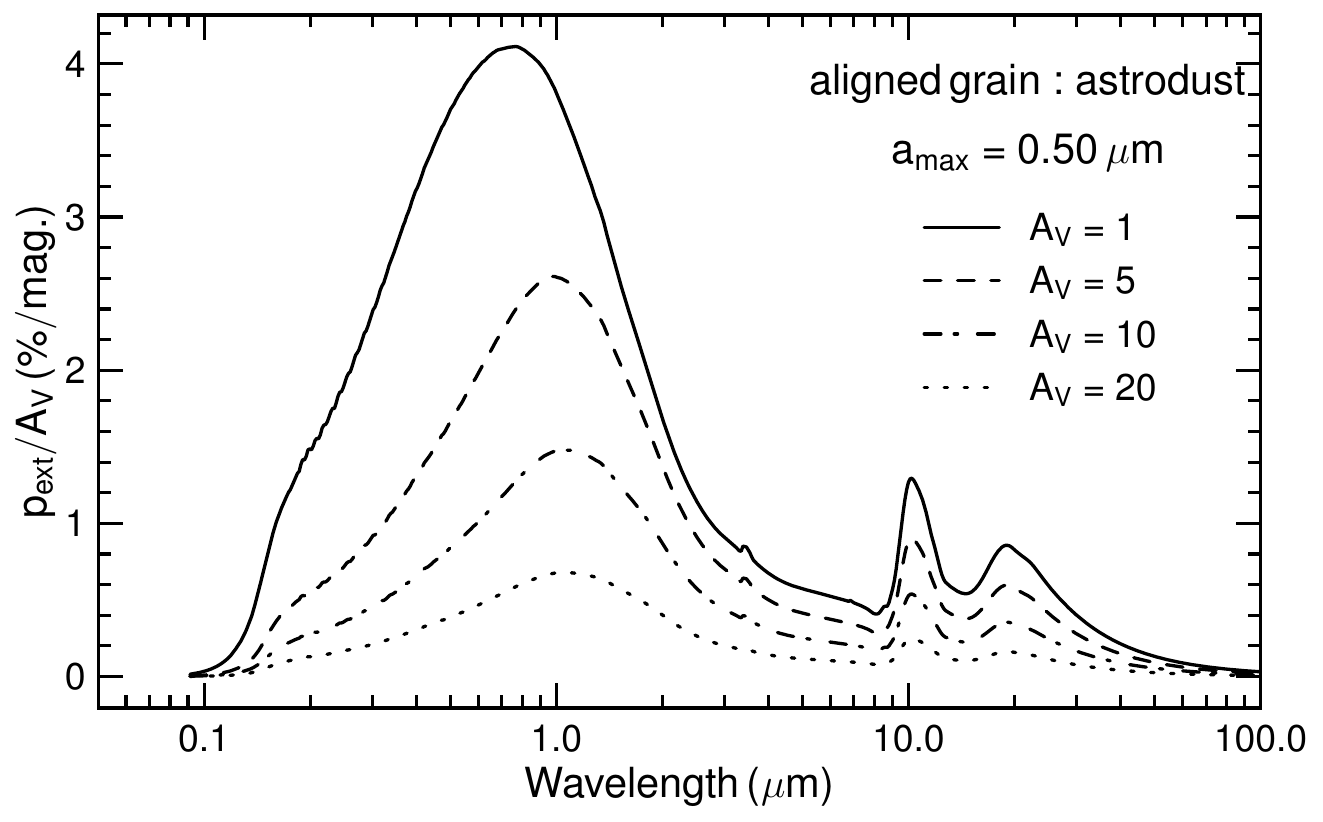}
    \caption{Polarisation spectrum $p_{\rm ext}/A_{\rm V}$ vs. $\lambda$ of starlight dust polarisation for different lines of sight ($A_{\rm V}$), shown for maximum grain sizes of 0.25$\,\mu$m (left panel) and 0.5$\,\mu$m (right panel). For $\lambda\leq 10\,\mu$m, the spectral feature is broader for larger maximum grain size. When $A_{\rm V}$ is low, smaller $a_{\rm max}$ yields a greater amplitude; however, this relationship inverts as $A_{\rm V}$ increases. The oblate grains with an axial ratio of 1.4 are adopted.}
    \label{fig:pol_spec_abs}
\end{figure*}
\begin{figure*}
\sidecaption
    \includegraphics[width=12cm]{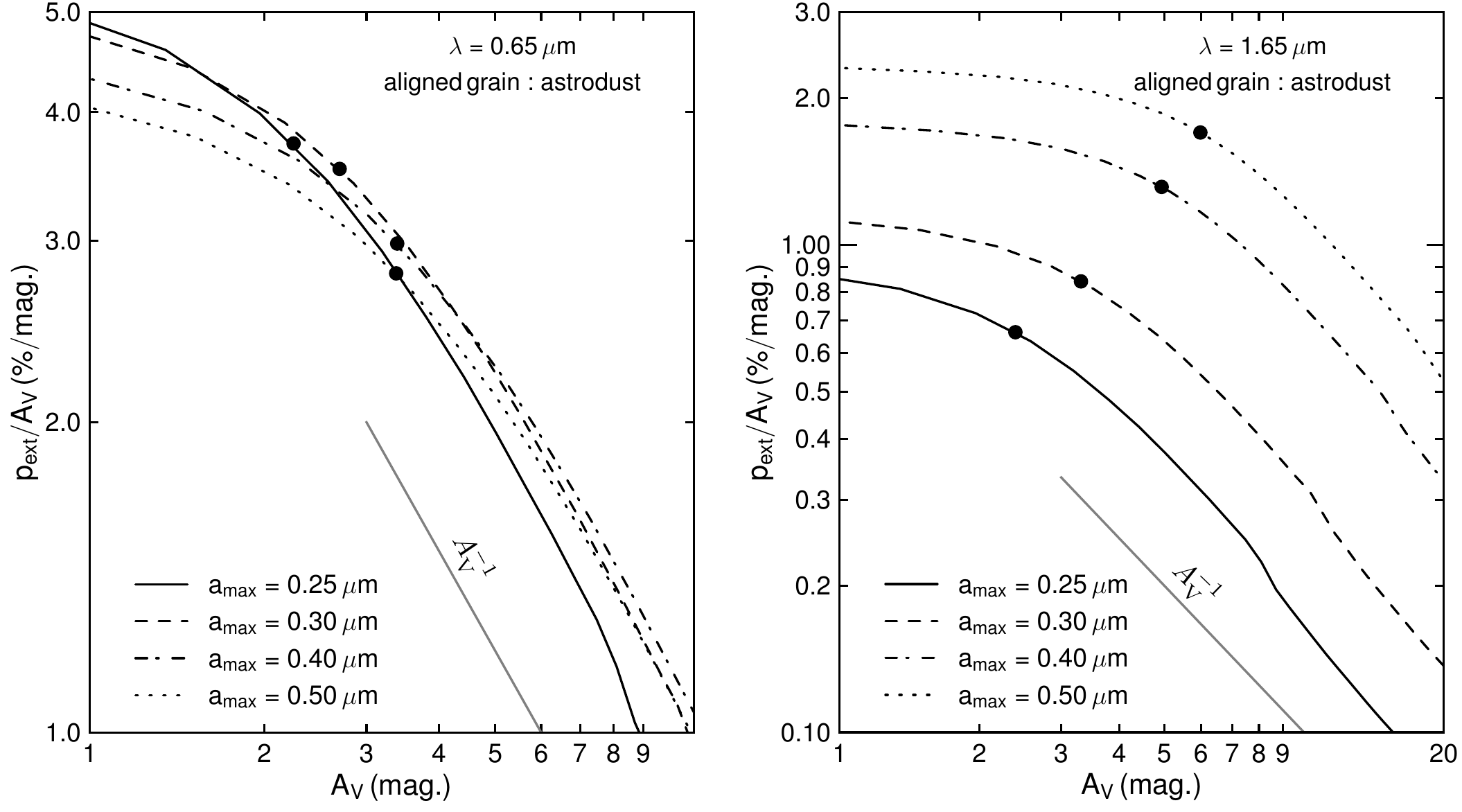}
       \caption{Relation between $p/A_{\rm V}$ and $A_{\rm V}$ for starlight polarisation at optical wavelength (R band, {\bf left panel}) and near-IR (H band, {\bf right panel}) wavelengths. Different lines represent different maximum grain sizes. The dots mark locations where the slope changes from $<-1$ to $\simeq -1$. A slope of $-1$ indicates the complete loss of grain alignment.}
        \label{fig:pAv_abs}
\end{figure*}
\begin{figure*}[!ht]
\sidecaption
    \includegraphics[width=12cm]{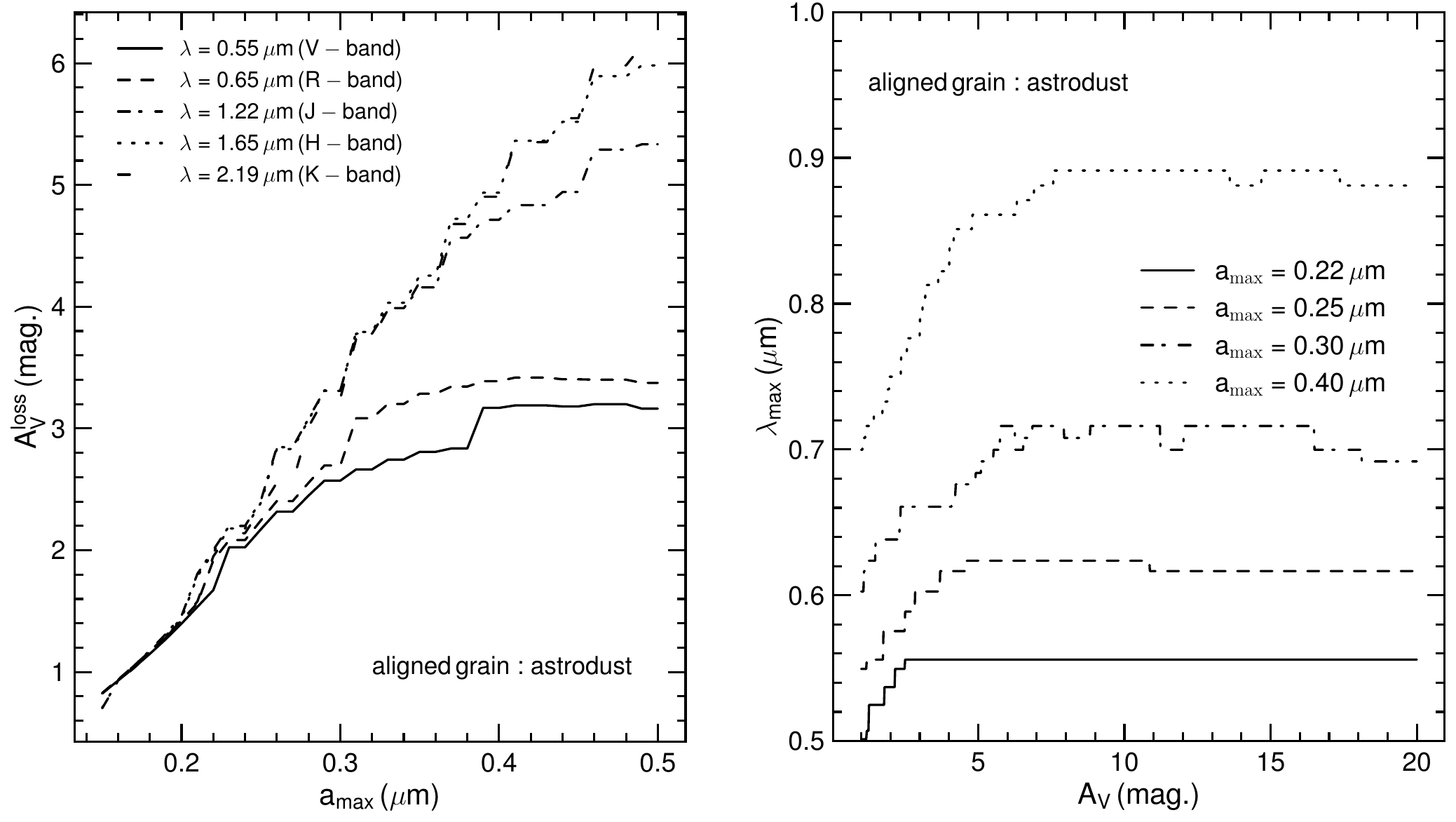}
        \caption{{\bf (Left panel)}: Relation between $A^{\rm loss}_{\rm V}$, where the grain completely loses its alignment, and maximum grain size. This plot is shown for several commonly used wavelengths from optical to near-IR. {\bf (Right panel)}: Relationship between the wavelength of peak starlight polarisation, $\lambda_{\rm max}$, and $A_{\rm V}$. The value of $\lambda_{\rm max}$ increases initially and then remains steady as $A_{\rm V}$ increases.}
        \label{fig:Avloss_amax_abs}
\end{figure*}

\begin{figure*}
    \centering
    \includegraphics[width=0.48\textwidth]{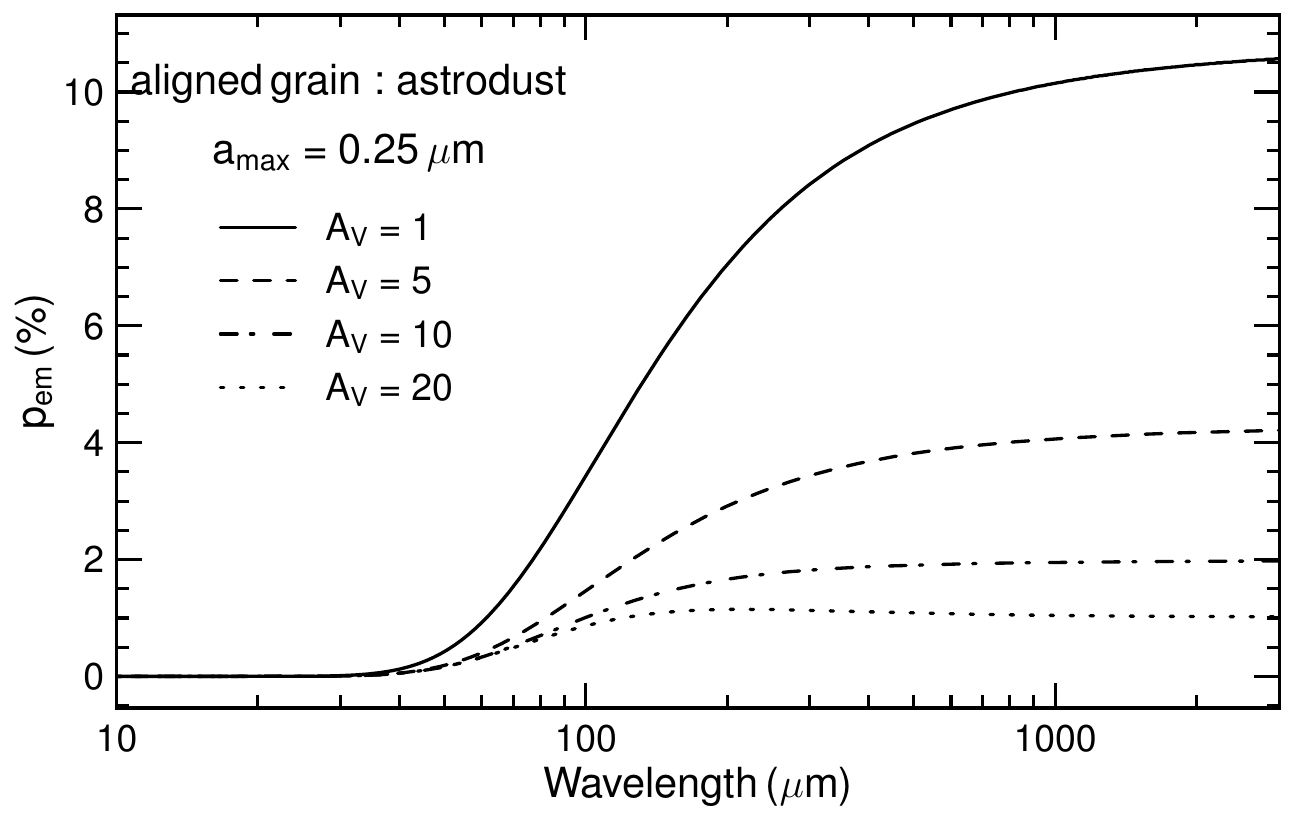}
    \includegraphics[width=0.48\textwidth]{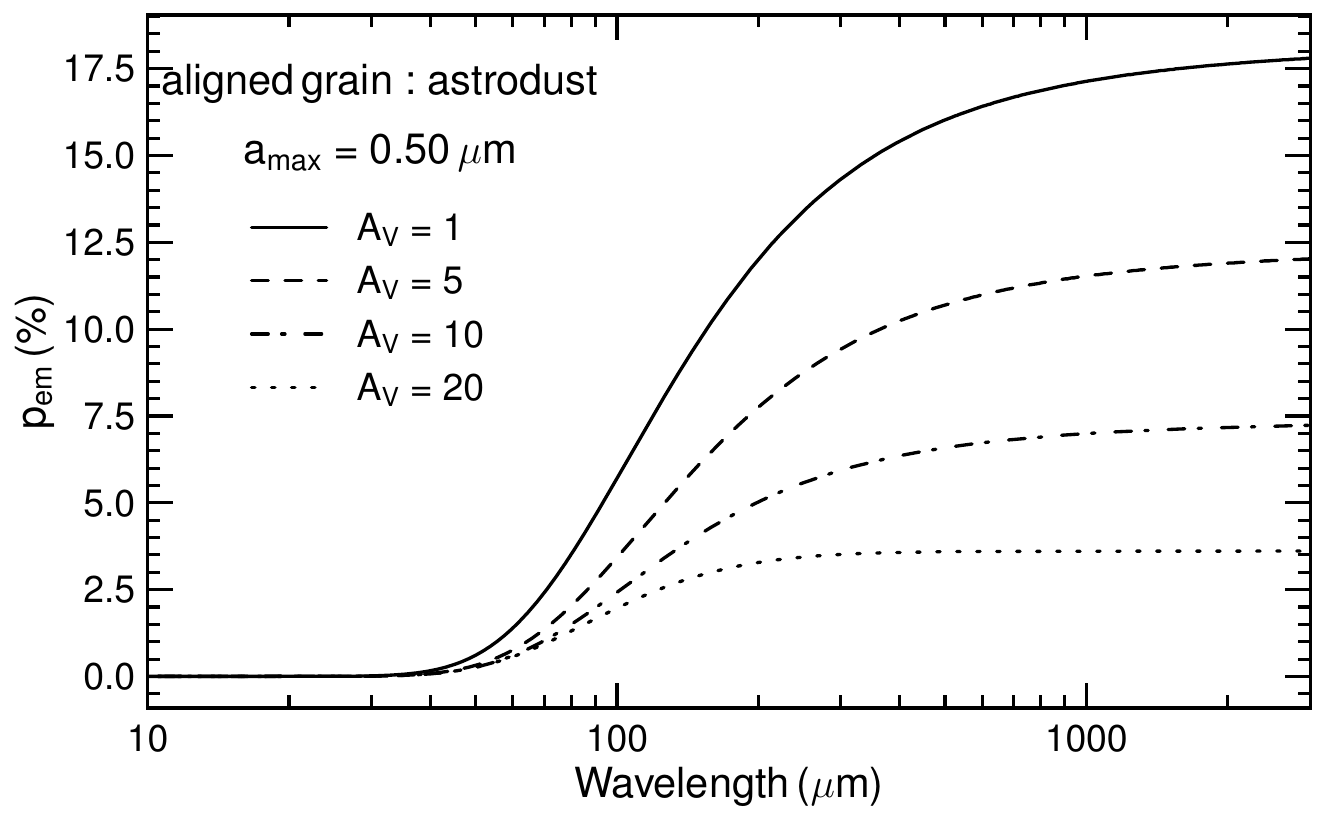}
    \caption{Polarisation spectrum of thermal dust polarisation towards a starless core for different values of the line-of-sight visual extinction ($A^{\rm los}_{\rm V}$), shown for $a_{\rm max}=0.25\,\mu$m ({\bf left panel}) and $a_{\rm max}=0.5\,\mu$m ({\bf right panel}). The spectrum first increases rapidly and then increases gradually towards longer wavelengths. Higher $a_{\rm max}$ yields a higher degree of polarisation under the same physical condition (characterised by $A_{\rm V}$).}
    \label{fig:pol_spec_emis}
\end{figure*}
\begin{figure*}
\sidecaption
    \includegraphics[width=12cm]{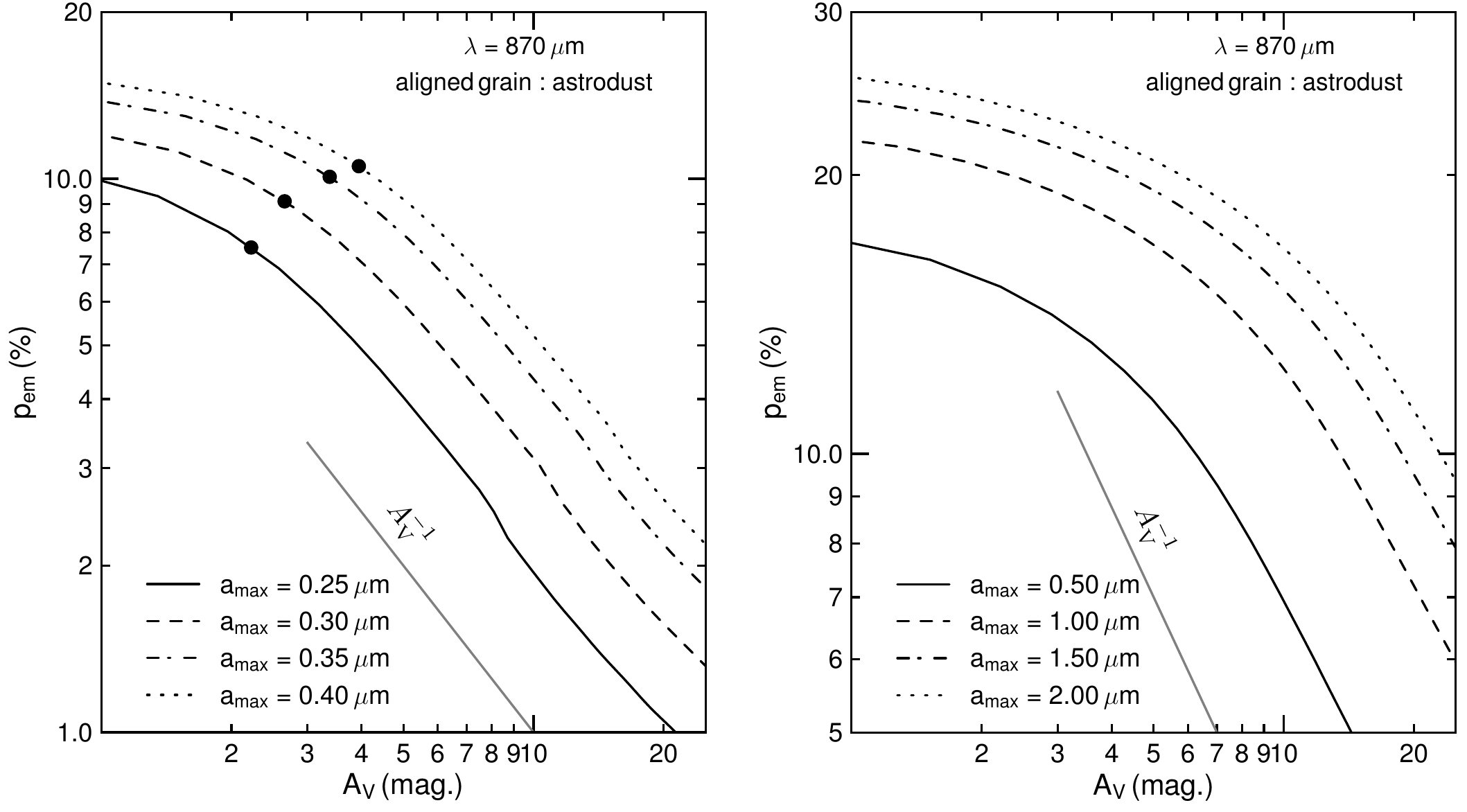}
    \caption{Same as in Figure \ref{fig:pAv_abs}. Relation between $p_{\rm em}$ and $A_{\rm V}$ at 870$\,\mu$m for different maximum grain sizes. The grey line shows a slope of $-1$, indicating a total loss of grain alignment. Smaller grains quickly lose their alignment within the cloud (left panel). If the grains are sufficiently large, the alignment is more efficient at the same $A_{\rm V}$, resulting in shallower slopes.} 
    \label{fig:PI_emis}

\end{figure*}
\subsubsection{Thermal dust polarisation}
In the presented version of our model, the degree of thermal dust polarisation was calculated for the optically thin case. With a uniform magnetic field orientation, the total and polarised intensity can be estimated analytically, as described by \cite{2005ApJ...631..361C,2020ApJ...896...44L, 2021ApJ...906..115T,HoangBao.2024,2024arXiv240317088T}. The total emission intensity at a given position along a line of sight (denoted by $r_{i}$ in Figure \ref{fig:sketch}) is calculated as
\begin{equation}
        dI_{\rm em} = n_{\rm H}ds \times \int^{a_{\rm max}}_{a_{\rm min}} Q_{\rm abs} \pi a^{2} 
        \times B_{\lambda}(T_{\rm d})\frac{1}{n_{\rm H}}\frac{dn}{da}da.~~~
\end{equation}
Naturally, this intensity depends on the line-of-sight $A_{\rm V}$, as determined in Equation \ref{eq:Av_obs}. The black-body radiation at dust temperature $T_{\rm d}$ is given by $B_{\lambda}(T_{\rm d})$, while $Q_{\rm abs}$ is the extinction coefficient and a function of $a$ and $\lambda$. In this study, we disregarded the distribution $T_{\rm d}$, which contrasts with the previous configuration in \textsc{DustPOL\_py}. This assumption is particularly justifiable for starless cores, where grains are probably large, making the nanodust contribution insignificant. The distribution of $T_{\rm d}$ for large grains is a delta function (see Figure 8 in \citealt{2020ApJ...896...44L}).

The local polarised intensity of thermal dust at the same location $r_{i}$ as above is given by
\begin{equation}\label{eq:Ipol_emi}
        dI_{\rm pol} = n_{\rm H}ds\times \int^{a_{\rm max}}_{a_{\rm align}} f(a)F_{\rm turb}\sin^{2}\psi Q^{\rm pol}_{\rm abs}\pi a^{2} \times B_{\lambda}(T_{\rm d})\frac{1}{n_{\rm H}}\frac{dn}{da}da, ~~~
\end{equation}
where $Q^{\rm pol}_{\rm abs}$ is the absorption polarisation efficiency, determined by $0.5\left[Q_{\rm abs}(\boldsymbol{E}\parallel \boldsymbol{a})-Q_{\rm abs}(\boldsymbol{E}\perp \boldsymbol{a})\right]$. These components were also taken from the {\textsc{astrodust} database}. The angle $\psi$ scales the amplitude of the degree of polarisation, but does not influence the shape of the spectrum, with the degree maximised for $\psi=90^{o}$ (i.e. when the magnetic field lies completely in the plane of the sky). As aforementioned, in the following, we therefore consider the case of $\psi=90^{o}$ and discuss its effect later in Section \ref{sec:Bfield_variation}. The degree of thermal dust polarisation along a given line of sight is computed as
\begin{equation} \label{eq:pem}
    p_{\rm em}(A_{\rm V}) = 100 \times \frac{\int_{s}dI_{\rm pol}}{\int_{s}dI_{\rm em}} ~~~{(\%)}.    
\end{equation}

\section{Numerical results} \label{sec:numerical_results}
\subsection{Physical properties}
We aim to compare the polarisation degree from observations of starless cores, such as Pipe-109, with a maximum $A_{\rm V}$ at a centre of 50$\,$mag. Accordingly, we adopt $n(\rm 0)=5\times 10^{5}\,\rm cm^{-3}$ at the centre, $r_{\rm flat}=0.024\,$pc, and $r_{\rm out}=0.6\,\rm pc$ for our starless core setup (see Section \ref{sec:pipe-109}); the density decreases towards the envelope as $r^{-2}$. The interstellar radiation field is taken as $U_{0}=1$ with $\gamma =0.3$ (for a dense core; see \citealt{2007ApJ...663.1055B}). 
For grain alignment, we consider the perfect alignment of large grains of size $a\gg a_{\rm align}$ with the magnetic field by the MRAT mechanism \citep{2016ApJ...831..159H} and set $f_{\rm max}=1$. The summary of the parameters is given in Table \ref{tab:parameters}.

We make the following key assumptions. Along an individual line of sight, while the physical profiles (gas density, temperature, radiation field, etc.) vary locally, we keep $a_{\rm max}$ constant along this line of sight, although $a_{\rm max}$ would not be the same for different locations along it.

The left panel of Figure \ref{fig:phys_prop} displays the map of visual extinction ($A_{\rm V}$) in the V band ($0.55\,\mu$m) along each line of sight in the plane of the sky ($\hat{x}\hat{y}$). Moving towards the centre, the density of the gas increases and the length $\hat{s}$ becomes longer, leading to a higher $A_{\rm V}$. The right panel of Figure \ref{fig:phys_prop} shows the corresponding alignment size map ($a_{\rm align}$) in the $\hat{x}\hat{z}$ plane. From the periphery to the centre, the gas density increases, while the radiation field notably decreases, and its mean wavelength becomes longer, resulting in a larger alignment size. Hence, to remain aligned deeper into the cloud, the average grain size must be larger.

\subsection{Starlight polarisation}
Figure \ref{fig:pol_spec_abs} shows the spectrum of starlight polarisation, $p_{\rm ext}/A_{\rm V}$, (the polarisation efficiency), as defined in Equation \ref{eq:pext} for different values of $A_{\rm V}$. The degree of polarisation decreases with depth into the core (higher $A_{\rm V}$). The polarisation is strongest in the optical to mid-IR, and then drastically drops at longer wavelengths. The left and right panels correspond to two different values of $a_{\rm max}$. Compared to the smaller $a_{\rm max}$, the larger grain size results in a lower maximum value of $p_{\rm ext}/A_{\rm V}$ at low $A_{\rm V}$ but a higher value at higher $A_{\rm V}$, and the degree of polarisation in the silicate absorption features at $\lambda=10,18\,\mu$m, is greater. Moreover, the spectrum is broader for larger $a_{\rm max}$. These effects arise from the contribution of larger aligned grains.

Figure \ref{fig:pAv_abs} illustrates the relationship between $p_{\rm ext}/A_{\rm V}$ and the line-of-sight $A_{\rm V}$ at two specific wavelengths: an optical wavelength (R band at 0.65$\,\mu$m, left panel) and a near-IR wavelength (H band at $1.65\,\mu$m, right panel). Generally, the polarisation efficiency decreases rapidly as $A_{\rm V}$ increases. Above a certain $A^{\rm loss}_{\rm V}$, $p_{\rm ext}/A_{\rm V} \sim A^{-1}_{\rm V}$, which implies the complete loss of grain alignment when $A_{\rm V}\geq A^{\rm loss}_{\rm V}$. The values of $A^{\rm loss}_{\rm V}$ marking the transition between shallow and steep slopes are marked by black dots\footnote{The values of $A^{\rm loss}_{\rm V}$ are determined by a piecewise linear fitting using the \textsc{pwlf}-library (\citealt{pwlf}). We use $-0.9$ as a criterion.}. This occurs because, towards higher $A_{\rm V}$, $a_{\rm align}$ increases, and the polarisation disappears where $a_{\rm align} > a_{\rm max}$. For a larger grain size that still maintains its alignment much deeper inside the cloud (higher $A_{\rm V}$), $A^{\rm loss}_{\rm V}$ is higher than for smaller sizes. Consequently, the polarisation in the optical band at the surface is higher for smaller grain sizes, as expected. Deeper inside the cloud, the alignment of smaller grains is rapidly reduced, whereas that of large grains remains. 

A similar trend is observed for the H band (right panel), but the alignment efficiency is lower for smaller $a_{\rm max}$ when the grain size is sufficiently small compared to the wavelength (that is, for $a\leq \bar{\lambda}/1.8$, the radiative torque decreases rapidly for smaller grains; see Equation \ref{eq:Q_gamma}). The left panel of Figure \ref{fig:Avloss_amax_abs} shows the dependence of $A^{\rm loss}_{\rm V}$ on $a_{\rm max}$ for several wavelengths in the optical and near-IR ranges, commonly used in observations. The value of $A^{\rm loss}_{\rm V}$ shows no significant change for smaller values of $a_{\rm max}$ but is affected when $a_{\rm max}$ is relatively larger. Longer-wavelength observations can detect dust polarisation at greater depths within the cloud.

Figure \ref{fig:pol_spec_abs} also shows that the wavelength $\lambda_{\rm max}$, where the polarisation spectrum reaches its maximum in the UV-NIR range, is longer for higher $A_{\rm V}$. The right panel of Figure \ref{fig:Avloss_amax_abs} illustrates the relationship between $\lambda_{\rm max}$ and $A_{\rm V}$. As $A_{\rm V}$ increases, $\lambda_{\rm max}$ initially increases and then becomes constant. The initial increase is due to an increase in the gas density. The values of $A_{\rm V}$ above which $\lambda_{\rm max}$ is saturated are higher for larger $a_{\rm max}$. Furthermore, $\lambda_{\rm max}$ is longer for larger $a_{\rm max}$. These features arise because larger grains can align more deeply within the cloud. The trend is consistent with previous studies in \cite{2008ApJ...674..304W,2015MNRAS.448.1178H,2020ApJ...905..157V}.

\subsection{Polarisation spectrum of thermal dust polarisation}
Figure \ref{fig:pol_spec_emis} demonstrates the polarisation spectrum for the thermal dust emission from Equation \ref{eq:pem} along different lines of sight. It initially exhibits an upward trend in the spectral profile, which then becomes constant at longer wavelengths. The spectra maintain their trend but vary in absolute values for different line-of-sight $A_{\rm V}$. The notable distinction is that larger grain sizes yield a greater degree of polarisation, $p_{\rm em}$, compared to smaller sizes under the same physical conditions.

Figure \ref{fig:PI_emis} shows the dependence of $p_{\rm em}$ on $A_{\rm V}$ at 870$\,\mu$m. This linkage, defined by the $p_{\rm em}-A_{\rm V}$ relation, reflects an alignment efficiency similar to that of $p_{\rm ext}/A_{\rm V}-A_{\rm V}$ for starlight, primarily because the $n_{\rm H}$ terms in the numerator and denominator of Equation \ref{eq:pem} cancel out, although $n_{\rm H}$ indirectly influences the alignment size. Typically, as the centre is approached, the degree of polarisation decreases approximately following the relation $p_{\rm em}\sim A^{-\eta}_{\rm V}$. Similarly to the starlight polarisation, this slope is shallower than $-1$ in the envelope and becomes steeper near the centre. The physical reason for this transition is the loss of alignment. Consequently, for larger grains ($a_{\rm max}>a_{\rm align}$), the slope is much shallower than $-1$. This critical $A_{\rm V}$ transition depends mainly on the maximum grain size ($a_{\rm max}$) and is independent of the grain composition and the maximum alignment factor $f_{\rm max}$. Interestingly, the slope can be slightly less than $-1$ in cases of no alignment due to integration over varying gas densities along the line of sight.

\section{Discussion} \label{sec:discussion}
\begin{figure*}
    \centering
    \includegraphics[width=0.95\textwidth]{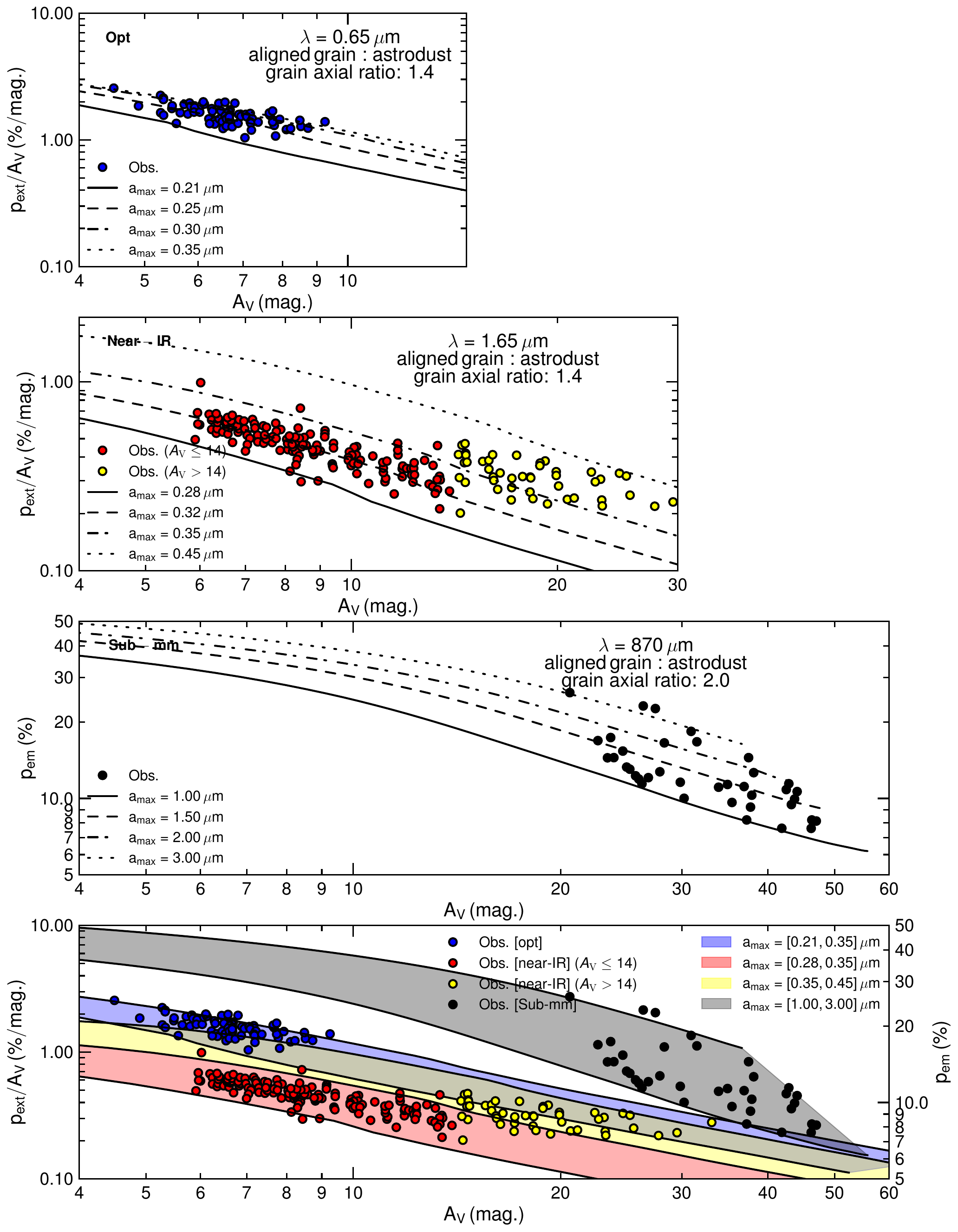}
    \caption{Comparison of $p_{\rm ext}/A_{\rm V}$ and $p_{\rm em}$ vs. $A_{\rm V}$ from our model and the data of Pipe-109. In the optical ({\bf first row}) and near-IR ({\bf second row}) wavelengths, we plot $p_{\rm ext}/A_{\rm V}$ (alignment efficiency) against $A_{\rm V}$, while for the submm wavelength ({\bf third row}), we use $p_{\rm em}$ versus $A_{\rm V}$. The fourth row combines all datasets above, with a shaded region indicating the range of $a_{\rm max}$ that covers the data shown in the above panels. The grain axial ratio at 870$\,\mu$m is set to 2, since it is too low for 1.4 shown in Figure \ref{fig:app_pipe-109-a_pav}. 
    Polarisation data are adapted from \cite{2014A&A...569L...1A} and references therein.}
    \label{fig:pipe-109_a_pav}
\end{figure*}
\begin{figure*}
    \centering
    \includegraphics[width=0.95\textwidth]{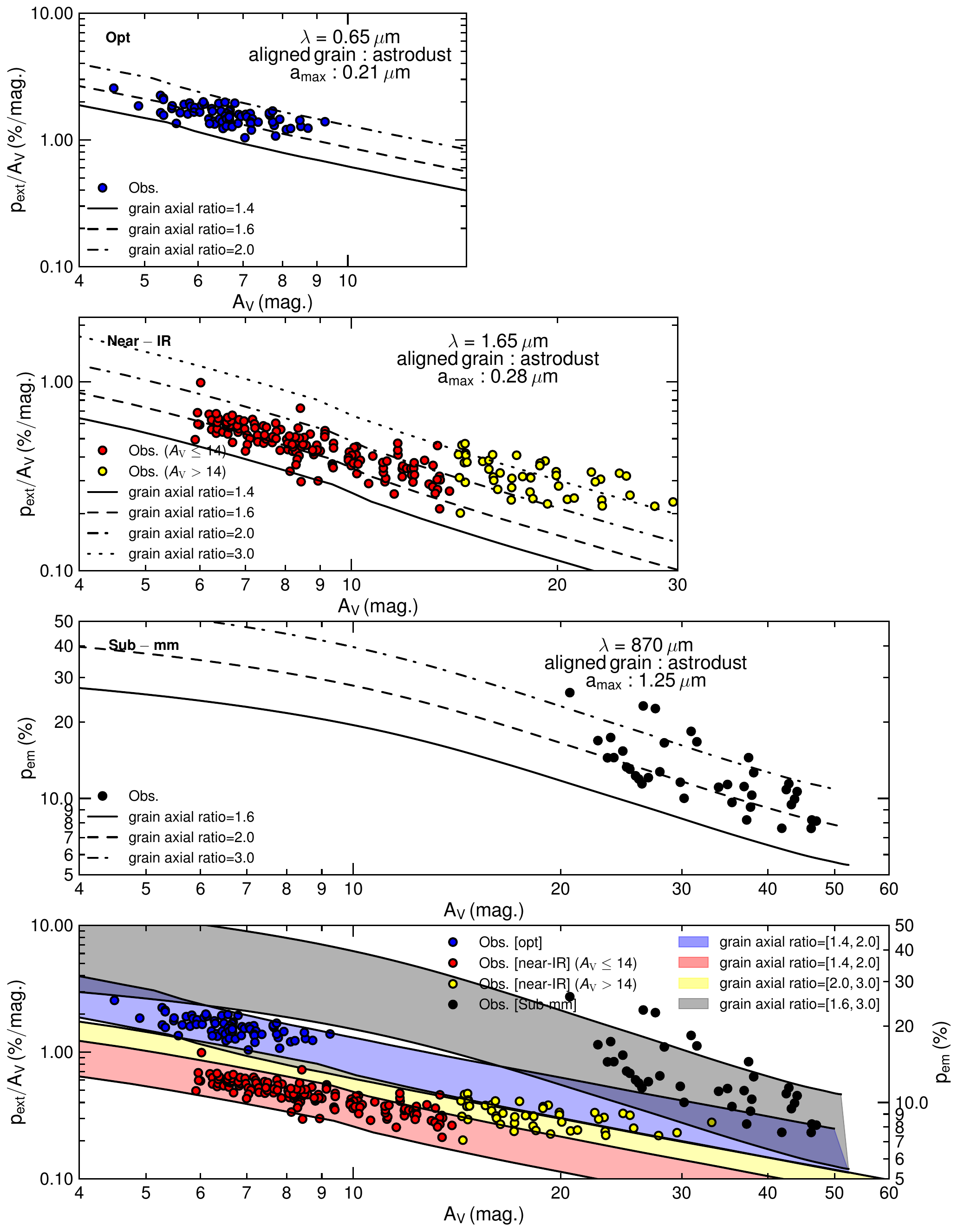}
    \caption{Similar to Figure \ref{fig:pipe-109_a_pav}, but here polarisation is calculated for different grain elongations, while keeping maximum grain sizes constant, namely 0.21, 0.28, and 1.25$\,\mu$m. For near-IR and submm data with $A_{\rm V}> 14\,$mag, even at an axial ratio of 3, the model fails to cover all data points, indicating the need for larger grains, as shown in Figure \ref{fig:pipe-109_a_pav}. For optical and near-IR data with $A_{\rm V}\leq 14\,$mag, the models fit well.}
    \label{fig:pipe-109_s_pav}
\end{figure*}
\subsection{Application to the Pipe-109 starless core} \label{sec:pipe-109}
We compare our modelling results with observational data from the specific starless core Pipe-109 (also known as FeSt 1-457). First, we adjust the parameters $r_{\rm flat}$ and $r_{\rm out}$ to match this specific target. To do so, we compare the line-of-sight visual extinction predicted by our model to that observed, as shown in Figure 1 of \cite{2014A&A...569L...1A}. The conditions are as follows: in the centre, $A_{\rm V} = 50\,\rm mag$, the diameter of $A_{\rm V}=20$ is approximately 0.1$\,$pc, while the area where $A_{\rm V}=10$ extends to about 0.2$\,$pc. The map of $A_{\rm V}$ in the left panel of the sky plane in Figure \ref{fig:phys_prop} (for $\alpha=2$ and $a_{\rm max}=1\,\mu$m) closely matches that of the Pipe-109, corresponding to $n_{0}=5\times 10^{5}\,\rm cm^{-3}$, $r_{\rm flat}=0.024\,$pc, and $r_{\rm out}=0.6\,$pc. 

\begin{figure*}
    \includegraphics[width=0.95\linewidth]{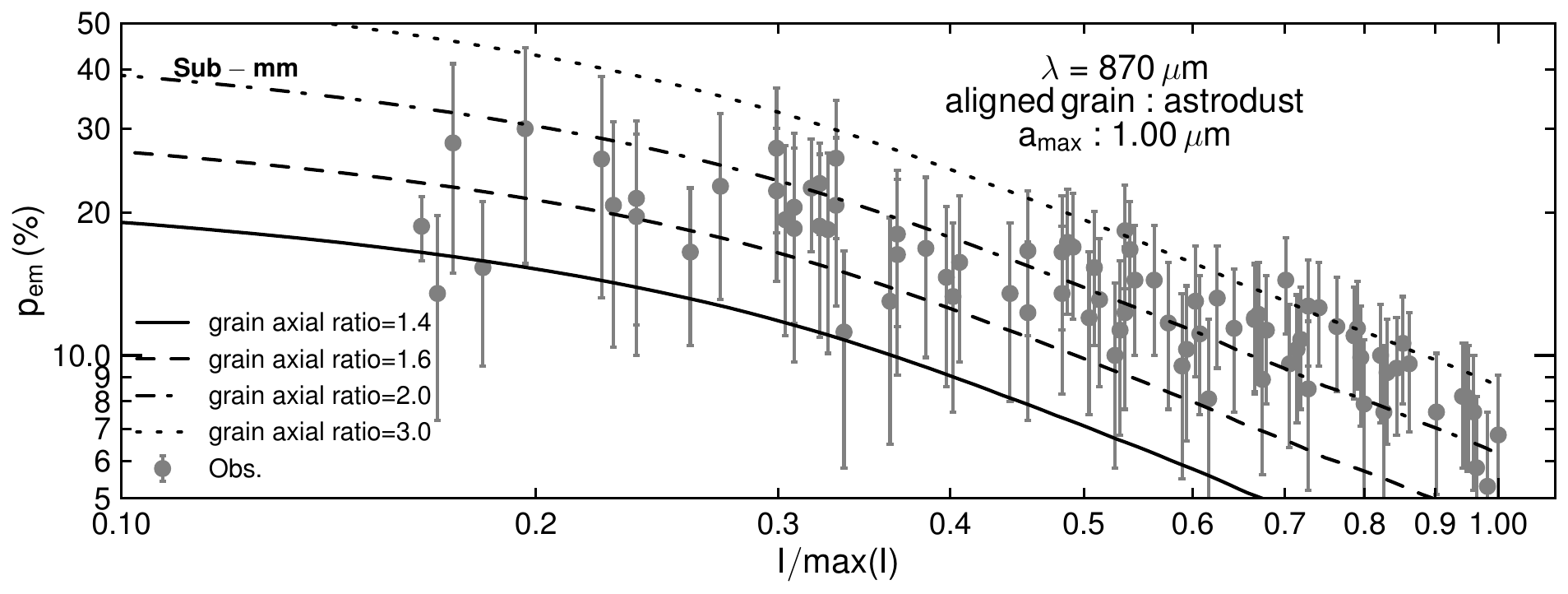}
    \caption{Similar to Figures \ref{fig:pipe-109_a_pav} and \ref{fig:pipe-109_s_pav}, but showing the $p_{\rm em}-I$ relationships for thermal dust polarization at 870$\,\mu$m. Our models nicely reproduce the observations. Data are taken from the public data file in \cite{2014A&A...569L...1A}.}
    \label{fig:pipe-109_pI}
\end{figure*}

Comparisons of $p_{\rm ext}/A_{\rm V}$ and $p_{\rm em}$ versus $A_{\rm V}$ between the predictions of our model, with a fixed grain elongation (an oblate grain with an axial ratio of 1.4 is used, except at 870$\,\mu$m, due to the fact that the predictions of our model are too low compared to the observed data, as illustrated in Figure \ref{fig:app_pipe-109-a_pav}), and the observations are shown in Figure \ref{fig:pipe-109_a_pav}, using data from \cite{2014A&A...569L...1A}. This comparison includes three wavelengths: 0.65$\,\mu$m (optical, first row), 1.65$\,\mu$m (near-IR, second row), and 870$\,\mu$m (submm, third row). In the model, only the maximum grain size, $a_{\rm max}$, is varied, as indicated by the different lines. To reproduce the data, $a_{\rm max}$ must be larger for dust polarisation at longer wavelengths. Specially, using the parameter sets mentioned above, our model predicts that the maximum grain size ranges from $0.21 \leq a_{\rm max} \leq 0.34\,\rm\mu m$ at a wavelength of $0.65\,\mu$m, $0.28 \leq a_{\rm max} \leq 0.45\,\rm \mu m$ at $1.65\,\mu$m, and $a_{\rm max} \geq 1\,\rm \mu m$ at $870\,\mu$m for a more elongated grain with the axial ratio greater than 1.4. For visualisation, the combination of all wavelengths is shown in the fourth panel, with the shaded areas corresponding to the optimal ranges of $a_{\rm max}$. The explicit values of $a_{\rm max}$ can change with different sets of parameters, but the requirement for larger $a_{\rm max}$ at higher $A_{\rm V}$ is unchanged. 

Two separte slopes are evident at 1.65$\,\mu$m, with one shallower for $A_{\rm V}\gtrsim 14\,$mag, which a single model cannot reproduce. However, distinct models with different values of $a_{\rm max}$ can reproduce this trend, with the shallower slope corresponding to a larger $a_{\rm max}$. In general, grain size increases towards the centre. Our finding may be consistent with the speculation in \cite{2014A&A...569L...1A}, in which the optical and near-IR wavelengths originate from the envelope of the starless core, while the submm observations originate from the core.

Figure \ref{fig:pipe-109_s_pav} shows the comparison with fixed $a_{\rm max}$ values of =0.21, 0.28 and $1.25\,\mu$m for the R, H, and submm bands (the lowest values in Figure \ref{fig:pipe-109_a_pav}) while varying the grain axial ratio. In this case, the agreement is poorer in the near-IR with $A_{\rm V}>14\,$mag and submm. Considering both the influence of the maximum grain size and elongation, reproducing these observations requires larger and more elongated grains deeper within the starless core.

Figure \ref{fig:pipe-109_pI} compares $p_{\rm em}-I$ at 870$\,\mu$m with a fixed maximum grain size of $1\,\mu$m for different grain elongations. Furthermore, this figure provides a direct comparison with Figure 3 of \cite{2014A&A...569L...1A}, in which the authors used the \textsc{DustPol} model from \cite{2012A&A...543A..16P}
and manually applied a fixed polarisation efficiency, setting it to 0.18 for gas densities up to $6\times 10^{4}\,\rm cm^{-3}$ and zero beyond that range. The \textsc{DustPol} model in \cite{2012A&A...543A..16P} does not account for the physics of grain alignment; instead, it treats alignment as an adjustable parameter ($\alpha$). Thus, our new physical modelling based on RAT physics provides a better approach to understanding dust polarisation in starless cores.

\subsection{Implications for grain alignment: magnetically enhanced radiative torque (MRAT)}
The successful fitting of the multi-wavelength polarisation data observed toward the starless core Pipe-109 with the RAT alignment model with a maximum alignment degree of $f_{\rm max} = 1$ and the composite \textsc{Astrodust} model suggests perfect alignment of large magnetic dust grains with $a\gg a_{\rm align}$. To achieve such a perfect alignment, the classical RAT theory alone is insufficient, which requires MRAT \citep{2008ApJ...676L..25L,2016ApJ...831..159H}. This represents the first evidence of MRAT in starless cores. Previously, evidence for MRAT was supported by various studies of the diffuse ISM \citep{Hensley.2023} and MCs \citep{Reissl.2020}. Recent dedicated models with appropriate treatments of internal and external alignment processes in \cite{2023MNRAS.520.3788G,2024MNRAS.530..984G,2024ApJ...970..114T,2024arXiv240710079C} show that MRAT with iron inclusions is required to understand the dust polarisations in protostars, low- and intermediate-mass young stellar objects, and protoplanetary disk observed by ALMA.

We note that a lower value of $f_{\rm max}$ may be sufficient (without requiring MRAT, i.e. classical RAT theory is adequate) if the grain elongation is increased to an axial ratio greater than 3, or if the radiation intensity is sufficient to compensate for the strong randomisation of grain alignment by dense gas collisions. The first case may be possible because we show that grain growth (increase in $a_{\rm max}$) is accompanied by increased grain elongation; however, such extremely elongated grains would not survive destruction mechanisms such as shattering or disruption (see \citealt{draine2024sensitivity}). The second case results in a higher degree of polarisation deeper within the cloud, making $p$ versus $A_{\rm V}$ relation shallower. However, there is no evidence of a nearby strong radiation source for the Pipe-109 core.

\subsection{Grains grow in starless cores}
Grain growth is expected to occur in dense clouds due to gas accretion and grain-grain collisions (see, e.g. \citealt{Hirashita.2013}), but the detailed physics of grain growth is not well understood. Traditionally, grain growth is assumed to be isotropic because of the random motion of gas and grains, which results only in an increase in grain size. However, a recent study by \citet{2022ApJ...928..102H} shows that grain growth becomes anisotropic due to grain alignment with ambient magnetic fields. As a result, grain growth can increase both in size and elongation, with larger grains having higher elongation.

In Section \ref{sec:pipe-109}, we show that the combination of starlight and thermal dust polarisation in starless cores clearly demonstrates the need for larger grains, which implies significant grain growth in such dense environments. To examine the extent of grain growth in different environments, sufficient observations of starless cores with distinct physical characteristics are needed, because the reduction in the degree of polarisation with total intensity or visual extinction differs for varying $a_{\rm max}$. Our proposed diagnostic for grain growth using the relation $p_{\rm ext}/A_{\rm V}$ versus $A_{\rm V}$, as shown in Section \ref{sec:numerical_results}, differs from and complements that previously described in \cite{2020ApJ...905..157V}, in which the authors inferred grain growth using the relation of the maximum wavelength $\lambda_{\rm max}$ with $A_{\rm V}$ from multi-wavelength observations in optical to near-IR (see their Figure 5 or our Figure \ref{fig:Avloss_amax_abs}, right panel).

Specially, Figures \ref{fig:pipe-109_a_pav} and \ref{fig:pipe-109_s_pav} show that higher elongation at the larger $A_{\rm V}$ can successfully reproduce the data. For instance, for maximum size of $a_{\rm max}\leq 0.28\mu$m, grain elongation between $1.4-2$ matches the starlight polarisation data for $A_{\rm V}<14$, but elongation of $\sim 2-3$ is required to fit the data at $A_{\rm V}\sim 10-30$. At $A_{\rm V}>30$, high elongation of $\sim 2-3$ and large grains are required to reproduce the upper bound of the thermal dust polarisation data. This provides the first evidence for grain growth accompanied by an increase in grain elongation, as predicted by the anisotropic grain growth model of \cite{2022ApJ...928..102H}. Another possible mechanism that produces more elongated grains is the centrifugal force arising from the fast rotation of grains induced by RATs \citep{Reissl.2024}. However, in starless cores, the radiation field is weak and cannot elongate grains deeper inside the cores. 

\subsection{Effects of magnetic field geometry} \label{sec:Bfield_variation}
Our numerical modelling assumed a uniform magnetic field in the plane of the sky. However, in realistic situations, the magnetic field varies along the line of sight.

First, magnetic turbulence can cause fluctuations of the local magnetic field with respect to the mean magnetic field. Due to this effect, the degree of polarisation is lower by a factor of $F_{\rm turb}<1$ (see Equations \ref{eq:pext} and \ref{eq:Ipol_emi}). Using the polarisation angles outlined in \cite{2014A&A...569L...1A} for the Pipe-109 core, and assuming that this cloud is sub-Alvénic, we computed turbulence factors $F_{\rm turb}$ of $0.98$, $0.96$, and $0.91$ for optical, Near-IR, and submm wavelengths, respectively. Evidently, the divergence from a uniform magnetic field scenario is minor at 2\%, 4\%, and 9\%, indicating that the influence of $F_{\rm turb}$ is insignificant for this specific instance of Pipe-109.

Second, the mean magnetic field may not lie in the plane of the sky, but may instead form an angle $\psi \neq 90^{o}$ with respect to the line of sight. Numerical simulations for the formation of cores by \cite{mocz2017moving} show that the inclination angle of the magnetic field varies from the outer to the inner region in cases of moderate and strong magnetic fields. Therefore, one cannot expect the magnetic field to be uniform, i.e. exhibit the same angle $\psi$ in starless cores. In this case, the degree of polarisation will be reduced by a factor of $f_{\rm max}\sin^{2}\psi$ (see Equations \ref{eq:pext} and \ref{eq:Ipol_emi}). For Pipe-109, \cite{kandori2018distortion} studied magnetic fields using near-IR and submm polarisation and report that the magnetic field can be described by a parabolic form. \cite{kandori2020distortion} applied the flux-freezing model from \cite{myers2018magnetic} to this core and find that the best-fit model yields the inclination angle of $\psi=35\pm 15^{\circ}$. Therefore, the model with a fixed grain elongation of $1.4$ could not reproduce the observational data with $a_{\rm max}$ mentioned above, because the maximum polarisation degree is reduced by a factor of $\sin^{2}35^{\circ}\approx 1/3$ compared to the ideal model of $\psi=90^{\circ}$. This suggests that the elongation of the grains must be enhanced to reproduce the observed polarisation.

\begin{figure}
    \centering
    \includegraphics[width=0.9\linewidth]{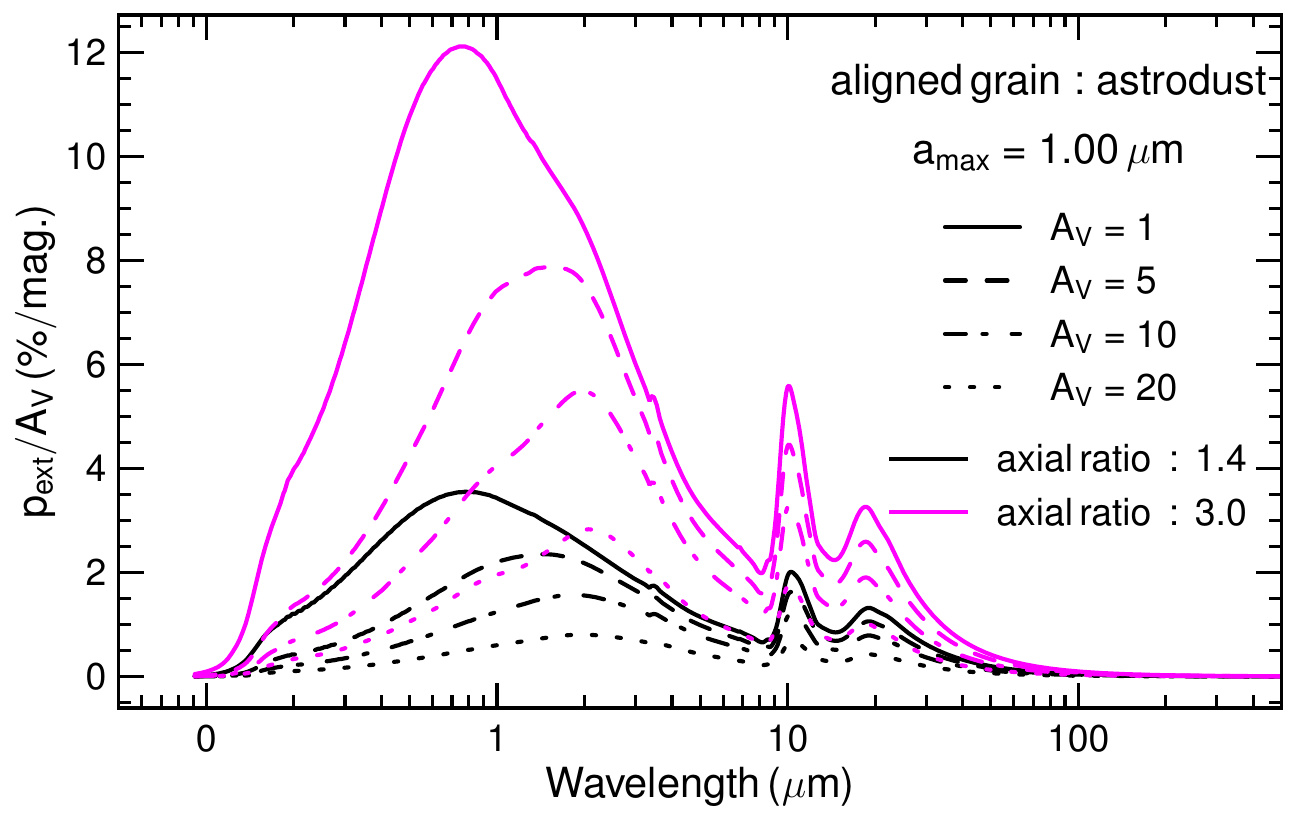}
    \includegraphics[width=0.9\linewidth]{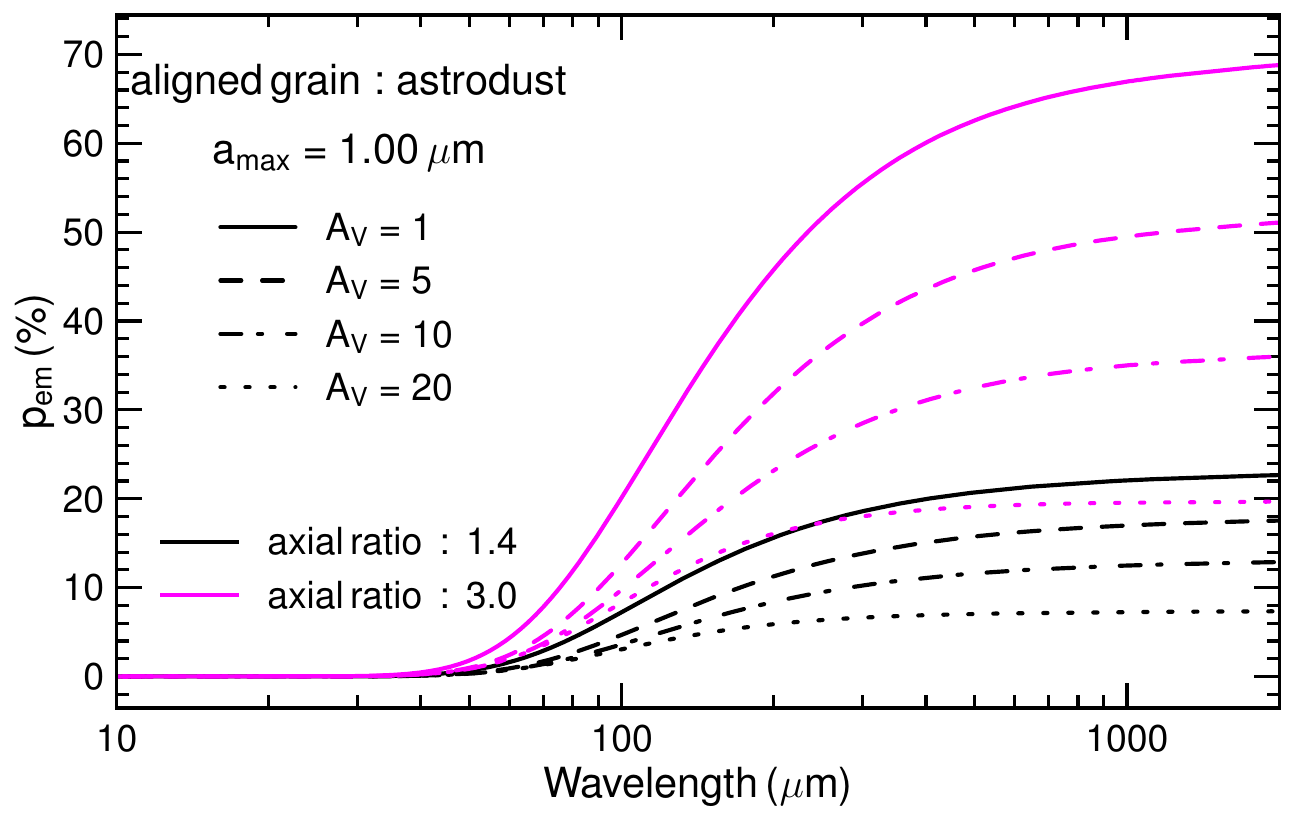}
    \caption{Polarisation spectrum in absorption ({\bf top panel}) and emission ({\bf bottom panel}) for a maximum grain size $a_{\rm max}=0.5\,\mu$m  with different axial ratios, of 1.4 and 3. The more elongated the grain, the higher the degree of polarisation.}
    \label{fig:grain_elongation}
\end{figure}

 To illustrate the effect of grain elongation on polarisation, Figure \ref{fig:grain_elongation} shows the polarisation spectrum of both starlight and thermal dust for oblate grains with axial ratios of 1.4 and 3. It is evident that more elongated grains result in higher polarisation. At the wavelengths discussed in this work ($0.65$, $1.65$ and $870\,\mu$m), the degree of polarisation for the axial ratio of 3 is  about 3.5 times that for the axial ratio of 1.4. Therefore, an accurate method to constrain the inclination angle of the magnetic field would help constrain grain elongation and grain growth physics \citep{2022ApJ...928..102H}. 

\subsection{Uncertainties}
The most obvious sources of uncertainty in this work are the assumption that the maximum grain size remains unchanged along the line of sight and that no treatment of grain evolution within the cloud is included. Proper models that consistently constrain the maximum grain size as a function of the physical condition could better reproduce the observations. However, as mentioned above, the quantitative value presented in this work might change, while the phenomena are expected to remain the same. The assumption that $T_{\rm gas} = T_{\rm dust}$ may not hold for $n_{\rm H}\leq 10^{6}\,\rm cm^{-3}$, as demonstrated in Figure 4 of \citealt{2019ApJ...884..176I}. The gas temperature can exceed that of the dust, potentially increasing the efficiency of rotational damping (see Equation \ref{eq:tau_damp}). Accurate modelling of heating and cooling processes could improve the accuracy of our model. In addition, the assumption of optical thickness for thermal dust polarisation holds at 870$\,\mu$m, but might cause uncertainty at shorter wavelengths (e.g. mid-IR to far-IR). 

\section{Conclusions} \label{sec:conclusion}
Significant advances have been made in our understanding of dust grain alignment physics, alongside a substantial increase in high-quality multi-wavelength dust polarisation data. Grain growth in dense clouds is the first step in planetesimal formation. However, the detailed mechanisms of grain alignment, as well as the process of grain growth and its underlying physics, are not well understood. Starless cores, which exhibit minimal internal dynamics and have the potential to form young stars, serve as the simplest and most suitable targets to better understand these physical processes. In this study, we aim to constrain the physics of grain alignment by radiative torques and dust evolution using the multi-wavelength dust polarisation observed toward starless cores. The main findings of our study are summarised as follows:

\begin{itemize}
    \item[1-] We model multi-wavelength polarisation for both starlight and emission using a newly proposed composition of interstellar dust (\textsc{astrodust}) and the MRAT alignment theory, with perfect alignment of large grains for a starless core in which grains are aligned by the attenuated interstellar radiation field. Different maximum grain sizes and elongations are considered, and the magnetic field is assumed to lie in the plane of the sky.

    \item[2-] Our model indicates that the reduction in the degree of polarisation towards the centre (referred to as the polarisation hole) is primarily caused by the loss of dust grain alignment of small grains in dense gas (due to gas collisional damping) and by the decrease in radiation intensity (attenuation). Larger grains can still maintain alignment deeper into the core due to the effect of longer wavelength photons.

    \item[3-] We tailor our model to a particular case of the Pipe-109 core as a showcase.
    \begin{itemize}
        \item The results of our optimal polarisation model (i.e., B-fields in the plane of the sky) can nicely reproduce the observational data consistently for optical (R band), near-IR (H band), and submm (870$\,\mu$m). 

        \item In particular, we find that grains must be larger and more elongated at higher visual extinction (deeper inside the cloud) to match observations. This is the first evidence for grain growth that involves both maximum size and grain elongation, consistent with the anisotropic grain growth model for aligned grains.

        \item When considering the magnetic field's inclination relative to the line of sight, the polarisation degree is noticeably reduced and can deviate significantly from observations. Therefore, the grain elongation must be increased to compensate for this reduction. 
    \end{itemize}
    
\item[4-] Finally, our model provides a promising tool for constraining the physics of grain alignment and grain evolution (GRADE-POL) using multi-wavelength dust polarisation.

\end{itemize}

\begin{acknowledgement}
The authors express gratitude to the anonymous referee for comments and corrections. The authors thank Dr. Isabelle Ristorcelli for useful comments from the early phase of this work. T.H. acknowledges the support from the main research project (No. 2025186902) from Korea Astronomy and Space Science (KASI). This work was partially supported by a grant from the Simons Foundation to IFIRSE, ICISE (916424, N.H.). The authors thank the ICISE staff for excellent support and hospitality.
\end{acknowledgement}

\bibliographystyle{aa} 
\bibliography{bib}

\appendix
\section{Models and observations in Pipe-109}
\begin{figure*}
\sidecaption
    \includegraphics[width=12cm]{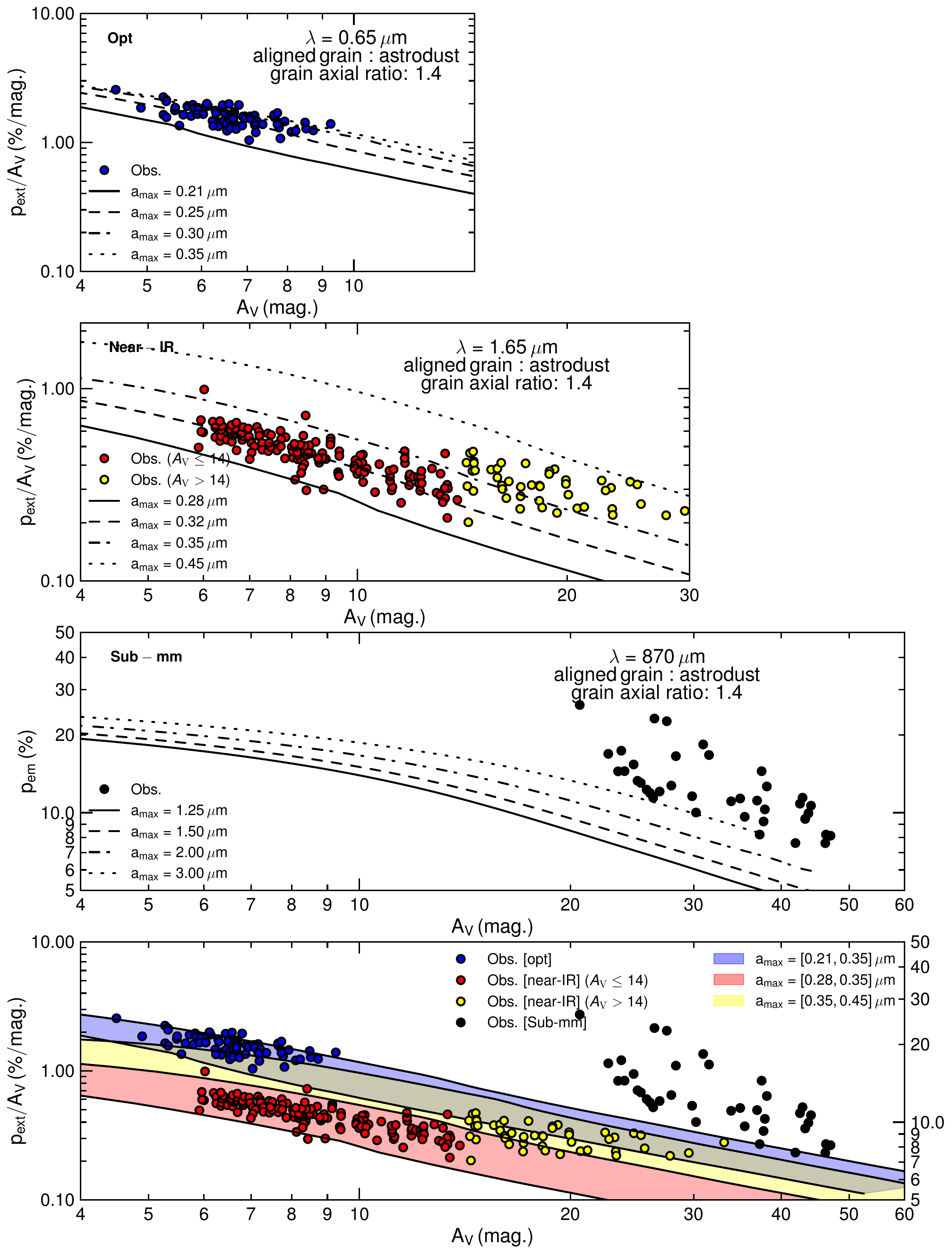}
    \caption{Similar to Figure \ref{fig:pipe-109_a_pav} but using the grains with an axial ratio of 1.4 for all bands. Our model is unable to reproduce the degree of thermal dust polarisation at 870$\,\mu$m.}
    \label{fig:app_pipe-109-a_pav}
\end{figure*}

Figure \ref{fig:app_pipe-109-a_pav} illustrates the comparison between our model and observations of Pipe-109, similar to Figure \ref{fig:pipe-109_a_pav}, when adopting the same grain elongation for the three bands. In the centre of the cloud, our model model fails to explain the observations at 870$\,\mu$m even using a relatively large grain of $3\,\mu$m. One requires more elongated grains. Therefore, we showed the thermal dust spectra predicted from our model with an axial ratio of 2, instead of 1.4 as for the starlight spectra in Figure \ref{fig:pipe-109_a_pav}.
\end{document}